\documentclass[12pt]{article}

\usepackage{textcomp}
\usepackage{chet}

\usepackage{amsmath}
\usepackage{amssymb}
\usepackage{graphicx}
\usepackage{caption}
\usepackage{subcaption}
\usepackage{float}
\usepackage{mdframed}
\usepackage{enumerate}
\usepackage{slashed}
\usepackage{tikz, tikz-3dplot, pgfplots}
\usepackage{tkz-graph}
\usetikzlibrary[positioning,patterns] 

\def\tr{\text{Tr}}
\def\CN{{\cal N}}

\title{Deformations of $W_{A,D,E}$ SCFTs}

\author{Ken Intriligator$^{}$\email{keni@ucsd.edu} and
Emily Nardoni$^{}$\email{enardoni@ucsd.edu}}

\affiliation{Department of Physics, University of California, San
Diego, La Jolla, CA 92093 USA}

\abstract{
We discuss aspects of theories with superpotentials given by Arnold's $A,D,E$ singularities, particularly the novelties that arise when the fields are matrices.   We focus on 4d ${\cal N}=1$ variants of susy QCD, with $U(N_c)$ or $SU(N_c)$ gauge group, $N_f$ fundamental flavors, and adjoint matter fields $X$ and $Y$ appearing in $W_{A,D,E}(X,Y)$ superpotentials.  Many of our considerations also apply in other possible contexts for matrix-variable $W_{A,D,E}$.  The 4d $W_{A,D,E}$ SQCD-type theories RG flow to superconformal field theories, and there are proposed duals in the literature for the $W_{A_k}$, $W_{D_k}$, and $W_{E_7}$ cases. 
As we review, the $W_{D_\text{even}}$ and $W_{E_7}$ duals rely on a conjectural, quantum truncation of the chiral ring.  We explore these issues by considering various  
deformations of the $W_{A,D,E}$ superpotentials, and the resulting RG flows and IR theories.  Rather than finding supporting evidence for the quantum truncation and $W_{D_\text{even}}$ and $W_{E_7}$ duals, we note some challenging evidence to the contrary.   
}

\date{April 2016} 

\begin{document}
\maketitle

\toc 

\newsec{Introduction}[Intro]

The simply-laced Lie groups, $A_k$, $D_k$, and $E_6$, $E_7$, and $E_8$ (``ADE'') relate to, and classify, far-flung things in physical mathematics. The Platonic solids are classified by the discrete subgroups $\Gamma _G\subset SU(2)$---cyclic, dihedral, tetrahedral, octahedral, and icosahedral---which connect to the ADE Lie algebras via the McKay correspondence\footnote{The irreducible representations $R_i$ of $\Gamma _G$ correspond to the nodes of the extended Dynkin diagram for $G$, with $R_F=\sum _j a_{ij}R_j$ for $R_F$ the fundamental of $SU(2)$ and $C_{ij}=2\delta _{ij}-a_{ij}$ the ADE Cartan matrix.}.  Another connection is in Arnold's simple surface singularities, which follow an ADE classification\rcite{Arnold}:
	\eqna{W_{A_k}&=X^{k+1}, \qquad W_{D_{k+2}}=X^{k+1}+XY^2, \\ W_{E_6}&=Y^3+X^4, \quad W_{E_7}=Y^3+YX^3,\quad  W_{E_8}=Y^3+X^5.}[WArnold]
These have resolutions, via lower order deformations, associated with the corresponding ADE Cartan, with the adjacency of the singularities that of the ADE Cartan matrix. 

In two dimensions, the ADE groups arise in the classification of minimal models and their partition functions \rcite{Cappelli:1986hf}.  The 2d $\CN =2$ minimal models with $\widehat c<1$ are given by 
 Landau-Ginzburg theories with the $W_{G=A,D,E}$ superpotentials \WArnold\ \rcite{Martinec:1988zu}, \rcite{Vafa:1988uu,Lerche:1989uy}. The chiral ring of the $W_{G}$ 2d $\CN=2$ SCFT is 
 related to the ADE group's Cartan, with $r_G=\text{rank}(G)$ chiral primary operators.  Deforming the theory by adding these chiral ring elements to the superpotential, $W\to W+\Delta W$, the deformation parameters can be associated with expectation values in the adjoint of $G$.  The deformation leads to multiple vacua, where the ADE group breaks into a subgroup. This breaking pattern is in accord with adjoint Higgsing, preserving the rank $r_G$ and corresponding to deleting a node from the extended Dynkin diagram, e.g.  
   \eqn{D_{k_1+k_2+2}\to D_{k_1+2}+A_{k_2}, \qquad E_7\to E_6+A_1, \qquad E_6\to D_5+A_1.}[adjointbreakingex]
The generic deformation gives $G\to r_G~A_1$, giving $\tr (-1)^F=r_G$ susy vacua. The solitons of the integrable $\Delta W$ deformations also exhibit the ADE structure, e.g. \rcite{Fendley:1992dm}.
 \begin{figure}[t]
\centering
	\begin{subfigure}[h]{0.76\textwidth}
		\includegraphics[width=\textwidth]{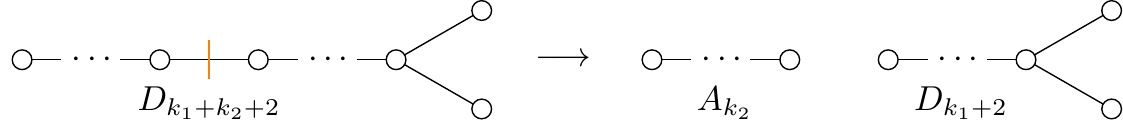}
		\label{}
	\end{subfigure}\\
	\begin{subfigure}[h]{0.81\textwidth}
		\includegraphics[width=\textwidth]{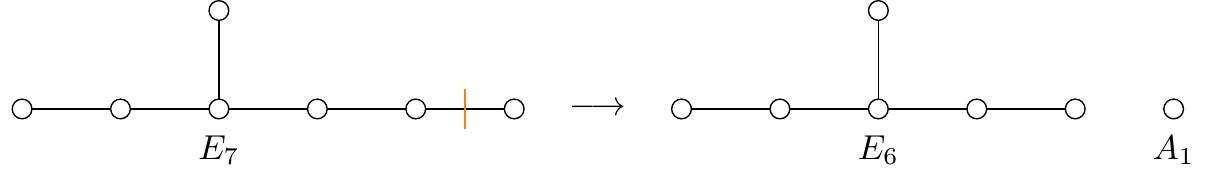}
		\label{}
	\end{subfigure}\\
		\begin{subfigure}[h]{0.68\textwidth}
		\includegraphics[width=\textwidth]{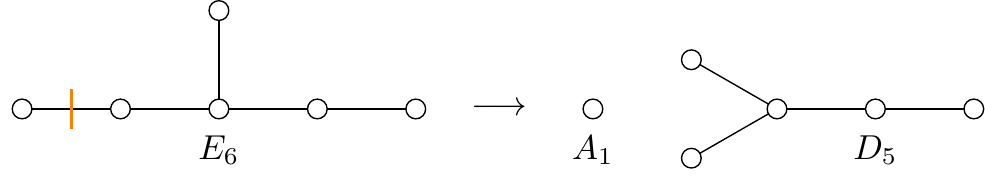}
		\label{}
	\end{subfigure}\\
\caption{Deforming the $W_{A,D,E}$ 2d $\CN=2$ SCFTs corresponds to adjoint Higgsing of the ADE group, hence cutting the Dynkin diagrams, as illustrated here for the flows in \adjointbreakingex. This gives the vacua associated with 1d representations of the $F$-terms.}\label{fig:ahiggsing}
\end{figure}

 A related connection with ADE groups is via local Calabi-Yau geometries: when the defining hypersurface has a singularity \WArnold, there are (collapsed) cycles corresponding to the ADE Dynkin diagram nodes, with intersections given by the group's Cartan matrix. String theory on these backgrounds can yield the corresponding ADE
gauge groups in spacetime \rcite{Witten:1995ex}. In this context, the geometric resolutions of the local singularities corresponding to $\Delta W$ deformations lead to adjoint Higgsing of the corresponding group.

\subsec{The chiral ring of $W_{A,D,E}(X, Y)$ for matrix fields $X$ and $Y$}[chiralringmatrix]

We are interested in an ADE classification that arises in the context of a family of 4d ${\cal N}=1$ SCFTs \rcite{twoadj}. Before delving into specifics, we highlight a difference in comparison with \WArnold: now $X$ and $Y$ are matrices, with 
	\eqna{W_{A_k}&=\tr (X^{k+1} +Y^2), \qquad W_{D_{k+2}}=\tr (X^{k+1}+XY^2), \\ W_{E_6}&=\tr(Y^3+X^4), \quad W_{E_7}=\tr(Y^3+YX^3),\quad  W_{E_8}=\tr(Y^3+X^5).}[matrixArnold]
The fact that matrices allow non-zero, nilpotent solutions to the equations of motion, and can have $[X,Y]\neq 0$, makes for important differences---even classically.

Recall that in theories with four supercharges, chiral primary operators have dimension proportional to their $U(1)_R$ charge, which is hence additive, and their OPEs yield the chiral ring. 
In terms of a microscopic, Lagrangian description, the chiral ring consists of gauge-invariant composites formed from the microscopic chiral superfields.  Superpotentials lead to chiral ring relations,  since $\partial _X W\sim \overline Q^2\partial _X K$ is not a primary, and is thus set to zero in the ring; for instance the LG theories \WArnold\ then have $r_{G=A,D,E}$ elements. 

We are here interested in aspects of the chiral rings for theories with matrix $X$ and $Y$ superpotentials \matrixArnold, and their $W\to W+\Delta W$ deformations.  Our focus is on the application to 4d ${\cal N}=1$ QFTs and renormalization group (RG) flows, but much of the analysis also applies to other possible contexts---for instance, in 2d or 3d--- where one could also consider theories with the superpotentials \matrixArnold\ with matrix fields. 

If the fields $X$ and $Y$ are $N_c\times N_c$ matrices, the superpotentials \matrixArnold\ have a $GL(N_c,C)$ symmetry under which $X$ and $Y$ transform in the adjoint representation: $X\to M^{-1}XM$, $Y\to M^{-1}YM$ for $M\in GL(N_c,C)$. Then, we can decide whether or not to gauge a subgroup of this symmetry, say $SU(N_c)$ or $U(N_c)$. If we do not gauge, \matrixArnold\ will leave unlifted a large space of flat directions. For instance, consider the matrix variable $A_k$ superpotential in  \matrixArnold, whose $F$-term chiral ring relations, $\partial _XW=\partial _YW=0$, are given by
	\eqn{W_{A_k}:   \qquad F_X\sim \partial _XW \sim X^k=0, \qquad F_Y\sim \partial _Y W\sim Y=0.}[Afterm]
$Y$ is massive and could be integrated out, setting $Y=0$; we merely included it here to make the ADE cases in \matrixArnold\ more uniform.  For $k=1$ and any $N_c$, $X$ is also massive, and there is a unique supersymmetric vacuum at $X=Y=0$.  For $k>1$ and $N_c=1$, \Afterm gives isolated vacua at $X=0$, and resolving the singularity by lower order $\Delta W$ shows that there are $\tr (-1)^F=r_G=k$ such vacua. For both $k>1$ and $N_c>1$, on the other hand, $X^k=0$ has a non-compact moduli space of flat direction solutions with nilpotent $X$; for example, $X$ could contain a block $v(\sigma ^1+i\sigma ^2)$ for arbitrary complex $v$. 

 In our context, $SU(N_c)$ or $U(N_c)$ is gauged, and the  nilpotent matrix solutions of \Afterm\ are lifted by the gauge $D$-term potential: supersymmetric vacua must have
	\eqn{V_D=0: \quad [X, X^\dagger]+[Y, Y^\dagger]+\hbox{other matter field contributions}=0.}[Dterms]
The ``other matter field contributions" are for example the contributions from $N_f$ fundamentals and anti-fundamentals $Q$, $\widetilde Q$ in variants of SQCD, which we need not consider for the moment; i.e. we consider the theory at $Q=\widetilde Q=0$.  For the $A_k$ case, \Dterms\
gives $[X, X^\dagger ]=0$, implying $X$ and $X^\dagger$ can be simultaneously diagonalized; then nilpotent solutions are eliminated, and \Afterm\ implies that the vacua are all at $X=0$.

The $D$ and $E$ cases, with $N_c>1$, have more matrix-related novelties since generally $[X,Y]\neq 0$.  For the $D$-series, the $F$-terms in the undeformed case are
	\eqn{W_{D_{k+2}}: \qquad F_X\sim X^k+Y^2=0, \qquad F_Y\sim \{X, Y\}=0.}[Dfterm]
The 1d representations are the same as in the $N_c=1$ case, giving $r_{D_{k+2}}=k+2$ chiral ring elements.  For matrices $X$ and $Y$, the chiral-ring relations \Dfterm\ lead to a qualitative difference between $k$ odd and $k$ even.  For $k$ odd, \Dfterm\ imply that $Y^3\sim YX^k\sim -YX^k=0$, and thus there are $3k$ independent chiral ring elements formed from $X$ and $Y$, given by 
	\eqn{k\ \text{odd}: \qquad \Theta _{\ell j}=X^{\ell-1}Y^{j -1},\qquad  \ell=1, \dots , k;\ \ j =1, 2,3.}[Dring]  
For $k$ even, $Y^{m\geq 3}\neq 0$ in the ring, so there are chiral ring elements with 
allowed values of $j$ that do not truncate, i.e. they do not have a maximum value independent of $N_c$.

Likewise, for $W_{E_6}$ the chiral ring relations 
	\eqn{W_{E_6}: \qquad F_X \sim X^3=0, \qquad F_Y\sim Y^2=0,}[Esixfterm]
allow for $r_{E_6}=6$ chiral ring elements with 1d representations, \{$1$, $X$, $Y$, $X^2$, $XY$, $X^2Y$\}.  For $N_c>1$, one can form, for example,  $\tr (XY)^\ell$ with arbitrary $\ell$ as independent chiral ring elements, so the ring does not truncate. Similarly, for $W_{E_7}$, the chiral ring relations
	\eqn{W_{E_7}:\qquad F_X\sim X^2Y+XYX+YX^2=0, \qquad F_Y\sim Y^2+X^3\sim 0,}[Esevenfterm]
lead to $r_{E_7}=7$ chiral ring elements when $N_c=1$, while for $N_c>1$ the classical chiral ring is not truncated. For $W_{E_8}$, the chiral ring relations
	\eqn{W_{E_8}:\qquad F_X\sim X^4=0, \qquad F_Y\sim Y^2=0,}[Eeightfterm]
lead to $r_{E_8}=8$ chiral ring elements for 1d representations ($X^{\ell-1}Y^{j -1}$ for $\ell=1,\dots , 4$ and $j =1,2$), but the classical chiral ring does not truncate for matrix representations.

\subsec{$W_{A,D,E}$ in 4d SQCD with fundamental plus adjoint matter}[4dcase]

We consider ADE superpotentials in the context of 4d ${\cal N}=1$ SCFTs, with gauge group $SU(N_c)$ or $U(N_c)$, $X$ and $Y$ adjoint chiral superfields, and $N_f$ (anti)fundamental flavors $Q$ (and $\tilde Q$). The possible interacting SCFTs were classified in \rcite{twoadj} as
	\eqn{W_{\widehat{O}} = 0, \qquad  W_{\widehat{A}} = \text{Tr} Y^2, \qquad W_{\widehat{D}} = \text{Tr} XY^2 , \qquad W_{\widehat{E}} = \text{Tr} Y^3}[WOADEhat]
along with \matrixArnold. The reappearance of Arnold's ADE classification in this context \rcite{twoadj} was unexpected.  Some interesting ideas and conjectures for a geometric explanation of the $W_{A,D,E}$ in this context appeared in \rcite{MR2399314}, in connection with matrix models and the construction of \rcite{MR2039033}.  We will not further explore these interesting ideas here. 

The IR phase of the theory depends on $N_f$ and $N_c$. It is convenient to consider these theories in the Veneziano limit of large $N_c$ and $N_f$, with the ratio 
	\eqn{x=N_c/N_f}[xdefis]
held fixed; the IR phase then only depends on $x$.  The $\widehat O$ theory is (or is not) asymptotically free for $x>1$ (or for $x\leq 1$), and RG flows to an interacting (or free electric) theory. 
 Larger $x$ values means that the theory is more asymptotically free, and hence the original ``electric" description is more strongly coupled in the IR.   The asymptotically free theories are expected\foot{ This can be seen e.g. for the $\widehat A$ theories with $N_f>0$ as in \rcite{Intriligator:1994sm}: a superpotential deformation leads to $\CN =2$ SQCD, and all the mutually non-local, massless monopole and dyon points in the moduli space collapse to the origin in the original theory. This has no free-field interpretation.}  to be in the interacting SCFT conformal phase for all $N_f<2N_c$ (i.e. $x>\frac{1}{2}$) for the $\widehat A$ cases, and for all $N_f<N_c$ (i.e. $x>1$) for the $\widehat O$, $\widehat D$ and $\widehat E$ cases.   For the $W_{A,D,E}$ theories \matrixArnold, on the other hand, there are more possible IR phases.  

In the $W_{A_1}$ case, the adjoints are massive and can be integrated out. The resulting IR theory is SQCD, which has the duality \rcite{Seiberg:1994pq}, with ``magnetic" gauge group $SU(N_f-N_c)$.  The dual reveals the bottom of the conformal window, and the existence of the IR-free magnetic phase for $\frac{2}{3}\leq x\leq 1$; for $x>1$, the theory generates a dynamical superpotential \rcite{Affleck:1983mk}. The $W_{A_{k>1}}$ theories were considered in \rcite{DKi,DKAS}, where a duality was proposed and checked.  Following \rcite{DKJLyqa,Kutasov:2014wwa} we write the $W_{A_k}$ duality in a way that will generalize to some cases:
	\eqn{\hbox{(some cases)}\quad W_G: \qquad SU(N_c) \leftrightarrow SU(\alpha _G N_f-N_c), \qquad\hbox{with}\quad \alpha _{A_k}=k.}[Adualalpha]
 Superpotential deformations of $W_{A_k}$ were considered in \rcite{DKNSAS}, where the fact that $\alpha_{A_k} =k$ was shown to tie in with the fact that upon a generic $\Delta W$ deformation, Arnold's $A_k$ singularity is resolved as
	\eqn{A_k\to k A_1,}[Akdef]
since the low-energy theory in each of the $k$ vacua has the $SU(n_i)\leftrightarrow SU(N_f-n_i)$ duality of \rcite{Seiberg:1994pq}.  The IR phases and relevance of the $W_{A_k}$ theories were clarified in \rcite{KPS} using $a$-maximization \rcite{Intriligator:2003jj}, including accounting for accidental symmetries.  

A duality of the form \Adualalpha\  for the case of two adjoint chiral superfields $X$ and $Y$, with $W_{D_{k+2}}$ as in \matrixArnold, was proposed in \rcite{JB}, with
	\eqn{\alpha _{D_{k+2}}=3k.}[Ddualalpha]
The IR phases and relevance of the superpotential terms were clarified in \rcite{twoadj}, where it was also noted how the $\alpha _{D_{k+2}}$ value \Ddualalpha\ can be understood / derived from $\Delta W$ deformations; this will be discussed much further, and clarified, in the present paper.

\begin{figure}[t]
	\centering
	\includegraphics[width=0.21\textwidth]{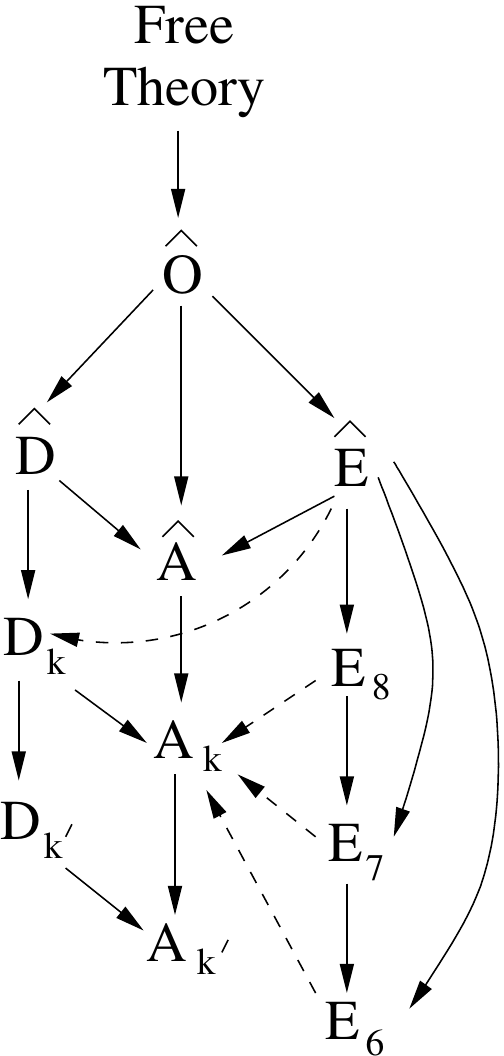}
	\caption{Flows among the fixed points of SQCD with two adjoints.\label{fig:RGflows}}
	\end{figure}

More recently, a duality for the case of $W_{E_7}$ was proposed in \rcite{DKJLyqa}, with
	\eqn{\alpha _{E_7}=30.}[Esevendualalpha]
The value \Esevendualalpha\ was moreover shown in \rcite{Kutasov:2014wwa} to be 
compatible with the superconformal index in the Veneziano limit\footnote{The exact matching of the electric and magnetic indices beyond this limit requires mathematical identities which have only been demonstrated explicitly for the $W_{A_1}$ SQCD duality case \rcite{Dolan:2008qi,Spiridonov:2009za}; the needed identities are conjectural for the $A_{k>1}$, $D_{k+2}$, and $E_7$ dualities.}, and it was argued \rcite{DKJLyqa,Kutasov:2014wwa} that the $W_{E_6}$ and $W_{E_8}$ theories cannot have duals of the simple form \Adualalpha; it is not yet know if these theories have duals. A motivating goal of our work was to obtain some additional insight into the value \Esevendualalpha, and its connection with the flows in Fig. \ref{fig:RGflows}.

\subsec{$W_{A,D,E}+\Delta W$ RG flows}[rgflowsintro]

Possible flows between these fixed points are illustrated in Figure \ref{fig:RGflows}, taken from \rcite{twoadj}. 

\noindent
We here emphasize that this figure is somewhat incomplete: the $\Delta W$ superpotential deformations give additional vacua, with a richer IR structure than is indicated in the figure.  Indeed, even the 1d ($N_c=1$) representations of the chiral ring of the deformed $W_{A,D,E}+\Delta W$ superpotentials give rank $r_{G=A,D,E}$ vacua, as in the examples \adjointbreakingex.  The two-matrix $D$ and $E$ cases with $N_c>1$ give additional vacua.  
Incidentally, much as in 2d, there are BPS solitons---here domain walls---interpolating between the vacua; we will not discuss them further here. 

To illustrate the multiple vacua and possibility for additional vacua, consider $W_{D_{k+2}}$ with $U(N_c)$ gauge group for $N_c>1$. The generic $\Delta W$ deformation gives\footnote{We use the standard notation for the floor and ceiling functions, $\lfloor x\rfloor$ and $\lceil x\rceil$, respectively.  So, for $k$ odd, $\lfloor \frac{1}{2}(k-1)\rfloor = \lceil \frac{1}{2}(k-1)\rceil = \frac{1}{2}(k-1)$; for $k$ even, $\lfloor \frac{1}{2}(k-1) \rfloor = \frac{1}{2}(k-2)$ and $\lceil \frac{1}{2}(k-1)\rceil = \frac{k}{2}$.}   \rcite{CachazoGH,twoadj}
	 \eqn{D_{k+2}\to (k+2)A_1^{1d}+\lfloor \frac{1}{2}(k-1)\rfloor A_1^{2d}.}[Dkdeform]
The $1d$ and $2d$ refers to the dimension of the representation of the (deformed) chiral ring. The higher-dimensional representations of the chiral ring are the new elements of the matrix-variable superpotentials.  The gauge group is then broken as \rcite{twoadj}
	\eqn{U(N_c)\to \prod_{i=1}^{k+2}U(n_i)\prod _{j=1}^{\lfloor \frac{1}{2} (k-1)\rfloor } U(n^{2d}_j)\qquad\hbox{with}\qquad \sum _{i=1}^{k+2}n_i +\sum _{j=1}^{\lfloor \frac{1}{2}(k-1)\rfloor }2n^{2d}_j=N_c.}[DkHiggsing]
For $k$ odd, the low-energy theory is SQCD for each factor, with $N_f$ flavors for the $U(n_i)$ groups and $2N_f$ flavors for the $U(n^{2d}_j)$ groups, and then the duality of  \rcite{Seiberg:1994pq} in each factor fits with the value \Ddualalpha\ \rcite{twoadj}.   

We will discuss even vs odd $D_{k+2}$ and the duality of \rcite{JB} in much more detail in what follows.  We will also report on our attempt to understand the duality \rcite{Kutasov:2014wwa}---and the value \Esevendualalpha---by considering various $\Delta W$ deformations, similar to \Dkdeform\ and \DkHiggsing.  

\subsec{$W_{A,D,E}$ flat direction flows}[fdflowsintro]

The $W_{A,D,E}$ theories can also be deformed by moving away from the origin, on the moduli space of supersymmetric vacua. There are fundamental matter flat directions associated with expectation values for the $Q$ and $\widetilde Q$ matter fields (e.g.  $\langle Q_{N_f}\rangle = \langle \widetilde Q_{N_f}\rangle \neq 0$), and adjoint flat directions associated with expectation values $\langle X\rangle$ and/or $\langle Y\rangle $, as well as mixed directions where both fundamentals and adjoints receive expectation values. We will here primarily focus on the purely adjoint flat directions.  

For $X$ and $Y$ adjoints of $SU(N_c)$ gauge group, there are certain flat directions which exist for special values of $N_c$ which do not exist for the $U(N_c)$ case.  For example, for $W_{A_k}$ there are flat directions when $N_c=kn$ for integer $n$; along such flat directions,
\eqn{SU(kn)\to U(n)^k/U(1),}[subreaking]
where in the low-energy theory each $U(n)$ factor is a decoupled copy of SQCD with $N_f$ flavors.  As we will review in Section \ref{sec:FDAk}, this gives another check of $\alpha _{A_k}= k$ in the duality \Adualalpha. We will discuss  similar checks of $\alpha _{D_{k+2}}=3k$, for the case of $k$ odd.  As we will emphasize, the $D_\text{even}$ case is quite different from $D_\text{odd}$; similar series of flat directions for $D_{\text{even}}$ and $E_7$ have a more subtle story.

For the cases where the classical chiral ring does not truncate---namely, $W_{D_{k+2}}$ for $k$ even and $W_{E_7}$---we show that there are classically unlifted flat directions given by matrix solutions to the $F$- and $D$-terms of the undeformed theories. We argue that these flat directions are not lifted or removed by any dynamics, and they thus present a possible challenge for the proposed duals for these theories.

\subsec{Outline}[outline]

The outline of the rest of this paper is as follows.  In Section \ref{sec:technicalreview} we review some technical details, including a review of the known and conjectured dualities for the 4d $W_{A,D,E}$ SCFTs, and a discussion of their moduli spaces of vacua---especially with respect to higher-dimensional vacua. In Section \ref{sec:oneadjsec} we review some aspects of the $W_{\hat{A}}$ and $W_{A_k}$ theories to set the stage for subsequent analysis. 

In Section \ref{sec:D}, we consider the $W_{D_{k+2}}$ theories: First, we study a matrix-related classical moduli space of supersymmetric vacua present for the $D_{\text{even}}$ theory, which poses a puzzle for duality for $D_{\text{even}}$, and argue that these flat directions are not lifted by quantum effects. We demonstrate that these flat directions seem to violate the $a$-theorem, and discuss possible resolutions to this puzzle. We then study $SU(N_c)$-specific flat directions of the $D_{k+2}$ theories, reviewing that such flat directions provide a nontrivial check  of the $D_{\text{odd}}$ duality, and then showing that they lead to puzzles for the conjectured $D_{\text{even}}$ duality. Next, we study RG flows from the $W_{D_{k+2}}$ SCFTs via relevant superpotential deformations, again finding nontrivial checks of duality for $D_{\text{odd}}$, and more hurdles for $D_{\text{even}}$. We conclude Section \ref{sec:D} with comments on hints as to how these puzzles might be resolved.

In Section \ref{sec:E}, we similarly analyze the $W_{E_7}$ SCFT. We study matrix-related flat directions and $SU(N_c)$-specific flat directions of the $E_7$ theory, which turn out to be analogous to the puzzling $D_{\text{even}}$ flat directions.  We then study some $\Delta W$ RG flows from the $W_{E_7}$ SCFT, noting some features in the resulting higher-dimensional vacuum structure that are new to the $E$-series. Finally, we conclude in Section \ref{sec:E68flows} with comments on future directions, and some discussion of how the present work might be applied to the $W_{E_6}$ and $W_{E_8}$ SCFTs. In an appendix, we explore additional $E$-series RG flows.

\newsec{Technical review}\label{sec:technicalreview}

\subsec{The $a$-theorem, and $a$-maximization}[atheoremsec]

The 4d $a$-theorem \rcite{Cardy:1988cwa,
Jack:1990eb,
Komargodski:2011vj} implies that the endpoints of all RG flows must satisfy
	\eqn{a_{UV}>a_{IR}.}[atheorem]
In superconformal theories, $a$ is related to the 't Hooft anomalies for the superconformal $U(1)_R$ symmetry as \rcite{Anselmi:1997am} (we rescale to a convenient normalization):
	\eqn{a(R)=3{\rm Tr} R^3-{\rm Tr}R.}[aRsusy]
In cases where $U(1)_R$ can mix with $U(1)_F$ global flavor symmetries, the exact superconformal R-symmetry is determined by $a$-maximization \rcite{Intriligator:2003jj}, by locally maximizing \aRsusy\ over all possible $U(1)_R$ symmetries.   Cases with accidental symmetries or irrelevant interactions require special care: one then maximizes \aRsusy\ over R-symmetries that are not obvious from the original description.  One situation where such enhanced symmetries are evident is when a gauge invariant operator saturates, or seemingly violates, an SCFT unitarity bound, e.g. for scalar chiral primary operators ${\cal O}$:
	\eqn{\Delta ({\cal O})=\frac{3}{2}R({\cal O})\geq 1.}[deltaR]
The inequality is saturated for free chiral superfields, and apparent violations instead actually saturate the inequality, with an accidental symmetry $U(1)_{\cal O}$ which only acts on the IR-free-field composite operator.  See \rcite{KPS} for how $a$-maximization is modified in such cases, and its application to the $\widehat A$ SCFTs.  See \rcite{twoadj} for additional applications to the other theories in Fig. \ref{fig:RGflows}, and additional discussion. 

The $a$-theorem \atheorem\ requires, for example, that $a$ decreases when a fundamental flavor is given a mass and integrated out,
	\eqn{a_{SCFT}(N_c, N_f)>a_{SCFT}(N_c, N_f-1),}[flavorreduced]
where SCFT refers to any of the SCFTs in Fig. \ref{fig:RGflows}.   In the  Veneziano limit, \flavorreduced\ for this RG flow requires (recall $x\equiv N_c/N_f$) \rcite{twoadj}
	\eqn{\frac{d}{dx}\left(x^{-2}a(x)/N_f^2\right)<0.}[aderivneg]
Upon computing $a(x)$ for the SCFTs in Fig \ref{fig:RGflows}, it is verified that $x^{-2}a(x)/N_f^2$ is indeed monotonically decreasing for small $x$, but then flattens out when $x$ is sufficiently large, e.g. at $x\approx 13.8$ for the $W_{E_6}$ SCFTs \rcite{twoadj}.  The $a$-theorem implies that some new dynamical effect must kick in for $x$ at or before the problematic range where \aderivneg\ is violated.

  One such effect, for sufficiently large $x$, is that a dynamical superpotential could be generated, and the theory is no longer conformal; this is referred to as the stability bound.  For $W_{A_k}$ theories, the stability bound is $x<x_{\rm stability}=k$ \rcite{Affleck:1983mk,DKi, DKNSAS}.  Another effect, which can occur for $x<x_{\rm stability}$, is that the theory could develop non-obvious accidental symmetries.   In cases with known duals, such accidental symmetries can be evident in the dual description, where it is seen that some superpotential terms---or the dual gauge interaction---become irrelevant when $x_{elec}$ is too large ($x_{mag}$ is too small).  It is satisfying that the $a$-theorem condition \aderivneg\ is indeed satisfied in the $W_{A_k}$ theories \rcite{KPS} and the $W_{D_{k+2}}$ theories \rcite{twoadj} upon taking such accidental symmetries into account.

\subsec{Duality for the 4d SCFTs}[matrix]

Recall that the chiral ring consists of gauge-invariant composites, e.g. meson, baryon, and glueball operators, formed from the microscopic chiral superfields: here $X$ and $Y$, the fundamentals and anti-fundamentals $Q$, $\widetilde Q$, and the gauge field strength fermionic chiral superfields $W_{\alpha}$, subject to classical and quantum relations.   Such theories,  with adjoint(s) $X$ (or $X$ and $Y$, or similarly, other two-index representations, e.g. in the examples in  \rcite{ILS}) only have a known dual if the chiral ring of products of the adjoint(s) {\it truncates}.  Here, {\it truncate} means that the number of independent elements in the ring is independent\footnote{There is a classical chiral ring relation that the adjoint-valued operator $X^{N_c}$ can be expressed in terms of products of lower powers $X^{\ell <N_c}$ and the $u_j\equiv \text{Tr} X^j$. To see this, write the characteristic polynomial $P(x, u_j)\equiv \det (x-X)=x^{N_c}-x^{N_c-1}u_1+\dots$, and note that $P(x, u_j)|_{x=X}=0$.  Thus one can write any gauge invariant $\text{Tr}X ^\ell ={\cal P}_\ell(u_1, \dots ,u_{N_c})$ for some polynomial ${\cal P}_\ell$.  As shown in \rcite{Cachazo:2002ry}, such relations can be modified by instantons for sufficiently large $\ell$.   See e.g. \rcite{Cachazo:2002ry,MazzucatoFE}, and references therein, for examples of chiral ring relations involving the adjoint-valued gaugino and gauge field chiral superfield $W_\alpha$, including the glueball operator $S\sim \text{Tr}W_\alpha W^\alpha$ and generalizations.  Relations involving $W_\alpha$ and $S$ will not be discussed in this current work.} of $N_c$. An example of an untruncated case is the $\widehat A$ theory, for which a basis of adjoint-valued products is given by
 $\Theta _j(X) = X^{j-1}$, for $j=1,\dots ,N_c$; such theories do not have a known dual.  A truncated case is $W_{A_k}$, where $\Theta _j(X)=X^{j-1}$, for $j=1,\dots ,k$.  

 More generally, suppose that a truncated case has a basis of elements $\Theta _j(X,Y)$, with $j=1,\dots,\alpha$; these are holomorphic products without traces, so gauge-invariant chiral ring elements are formed by taking traces or contracting with $Q$ and $\tilde Q$.  One can form dressed quarks $Q_{(j)}\equiv \Theta_j(X,Y) Q$, which can then be used to construct gauge-invariant operators, such as the $\alpha N_f^2$ mesonic operators
	\eqn{M_j=\widetilde Q \Theta _j Q, \qquad j=1, \dots, \alpha.}[mesonsj]
(We suppress flavor indices: each $M_{j}=(M_{j})_{f, \widetilde f}$ is in the $(N_f, N_f)$ of $SU(N_f)_L\times SU(N_f)_R$.  For $SU(N_c)$ there are also baryonic operators, built out of the dressed quarks:
	\eqn{ 
	B^{(l_1,...,l_\alpha)} = Q_{(1)}^{l_1}...Q_{(\alpha)}^{l_\alpha},\ \ \ \sum_{j=1}^\alpha l_j = N_c.
	}[baryons]

As shown in \rcite{DKJLyqa}, the many constraints on any possible dual---including matching of the chiral operators, invariance under the same global symmetries, 't Hooft anomaly matching, and matching of the superconformal index \rcite{Kutasov:2014wwa}---essentially determines the dual (assuming it is of a similar form) to have gauge group $SU(\tilde{N}_c)$, with $\tilde{N}_c = \alpha N_f-N_c$, again with $N_f$ flavors $q$ and $\widetilde q$ in the (anti)fundamental of the gauge group, and adjoint fields we denote by $\hat{X},\hat{Y}$. The ratio \xdefis\ of the dual theory is 
	\eqn{\hat{x} = \tilde {N}_c/N_f=\alpha - x.}[hatx]  
The electric mesons \mesonsj\ map to elementary operators of the dual theory, which couple in $W_{\rm dual}$ to a corresponding mesonic composite operator in the magnetic theory.  Magnetic baryons map to electric baryons as 
	\eqn{
	B^{(l_1,...,l_\alpha)} \leftrightarrow \hat{B}^{(\hat{l}_1,...,\hat{l}_\alpha)},\ \ \ \hat{l}_j = \alpha N_f - l_j.
	}[dualbaryons]
The truncation of the ring to $\alpha$ generators  is a necessary ingredient for these classes of conjectured dualities. The chiral ring of the electric theory truncates classically in the $A_k$ and $D_\text{odd}$ cases, and has been conjectured to truncate quantum mechanically in the $D_\text{even}$ \rcite{JB} and $E_7$ \rcite{DKJLyqa} cases. 

The $W_{A,D,E}$ theories are understood in terms of the RG flows in Fig. \ref{fig:RGflows}, starting from the top, $W=0$ theories. 
   If $x>1$ the gauge coupling is asymptotically free, so it is a relevant deformation of the UV-free fixed point, driving the RG flow of the top arrow in Fig. \ref{fig:RGflows} into the $\widehat O$ SCFT.   Deforming by $W_{\widehat A}$, $W_{\widehat D}$, or $W_{\widehat E}$ gives flows, as in the figure, that are also all relevant for $x>1$ (the $\widehat A$ case can be defined down to $x>\frac{1}{2}$).  Generally, as long as the gauge coupling is asymptotically free, its negative contribution to anomalous dimensions drives the cubic superpotential terms to be relevant.  The $\widehat A\to A_k$, and $\widehat D\to D_{k+2}$, and $\widehat E\to E_r$ flows with non-cubic terms in $W(X,Y)$  only occur 
    if $x>x_{\text{min}}$, such that the negative anomalous dimension from the gauge interactions is large enough to drive the $W(X,Y)$ terms relevant; the values of $x_{\text{min}}$ were obtained using $a$-maximization for $W_{A_k}$ in \rcite{KPS} and in \rcite{twoadj} for the other $W_{G=A,D,E}$ theories.  Duality, if it is known and applicable, clarifies the IR phase structure of the theories for $x>x_\text{min}$, where the magnetic dual becomes more weakly coupled.  The fixed point theories whose duals are known or conjectured all have a similar phase structure \rcite{KPS,twoadj,DKJLyqa}:

\begin{center}
\begin{tabular}{c l}
$x\leq 1$ & free electric \\
$1 < x \leq  x_{\text{min}}$ & $(\widehat{A},\widehat{D},\widehat{E})$ electric \\
$x_{\text{min}} < x < \alpha - \hat{x}_{\text{min}}$ & $(A_k, D_{k+2}, E_r)$ conformal window \\
$\alpha - \hat{x}_{\text{min}} \leq  x < \alpha - 1$ & $(\widehat{A},\widehat{D},\widehat{E})$ magnetic \\
$\alpha - 1 \leq  x \leq \alpha $ & free magnetic \\
$\alpha <x $ & no vacuum 
\end{tabular}
\end{center}

\subsec{Moduli spaces of vacua of the theories}[moduli]

Recall that 4d $\CN =1$ theories with $W=0$ have a classical moduli space of vacua ${\cal M}_{cl}$, given by expectation values of the microscopic matter fields, subject to the $D$-term conditions \Dterms and modulo gauge equivalence.  Alternatively, ${\cal M}_{cl}$ is given by expectation values of gauge invariant composite, chiral superfield operators, modulo classical chiral ring relations (see for instance \rcite{Luty:1995sd}).  When $W\neq 0$, one also imposes the $F$-term chiral ring relations.  The quantum moduli space ${\cal M}_{qu}$ can be (fully or partially) lifted if $W_{dyn}$ is generated, or deformed for a specific $N_f$ as in \rcite{Seiberg:1994bz} or variants\foot{There are exotic examples of classical flat directions that are lifted by, for example, confinement (see e.g. \rcite{Intriligator:1996pu}); this can only occur if a gauge group remains unbroken and strong there.}; the constraints of symmetries and holomorphy often exactly determine the form of such effects, and with sufficient matter (e.g. sufficiently small $x$) this implies that $W_{dyn}=0$ and ${\cal M}_{cl}\cong {\cal M}_{qu}$.  

We will here focus on vacua with $Q=\tilde Q=0$, with non-zero expectation values for the adjoints, $X$ and $Y$; such vacua preserve the $SU(N_f)_L\times SU(N_f)_R$ global flavor symmetry.  
The $N_c\times N_c$ matrices $X$ and $Y$ are decomposed into  multiple copies of a set of basic, irreducible solutions of the $D$- and $F$-flatness conditions. We refer to such a basic vacuum solution representation as being $d$-dimensional if $X$ and $Y$ are represented as $d\times d$ matrices, which cannot be decomposed into smaller matrices.  

For the $\widehat A$ and $A_k$ theories and their $\Delta W$ deformations, we can set $Y=0$ and the $D$-terms give $[X,X^\dagger]=0$. Thus, $X$ and $X^\dagger$ can be simultaneously diagonalized by an appropriate gauge choice, and all vacuum solutions are $d=1$ dimensional, represented by eigenvalues on the diagonal of $X$. 
More generally, vacua with $[X,Y]=0$ allow for simultaneously diagonalizing $X$, $X^\dagger$, $Y$, and $Y^\dagger$, so the representations are $d=1$ dimensional.  
For cases other than $\widehat A$ and $A_k$ in Fig. \ref{fig:RGflows}, there are generally also 
 $d>1$ dimensional vacua, where $[X, Y]\neq 0$.  In such cases, 
we cannot in general fully diagonalize neither $X$ nor $Y$.  We can use the gauge freedom to e.g. diagonalize the real part of $X$ (or $Y$), and then impose the $D$-term to get an adjoint-worth of constraints on the remaining three real adjoints. We indeed find examples of vacua where neither $X$ nor $Y$ can be fully diagonalized.

The independent representations for $X$ and $Y$ vacuum solutions can be characterized by the independent solutions for the Casimir\footnote{Casimir here means matrices commuting with $X$ and $Y$, not the $U(N_c)$ or $SU(N_c)$ Casimir traces.} products of $X$ and $Y$. For example, if the $F$-terms imply that $[X^3,Y]=0,\ [Y^2,X]=0$ then we use the eigenvalues $X^3=x^3{\bf 1}_d,\ Y^2=y^2 {\bf 1}_d$ to label the vacua. In some cases we find there are no such Casimirs (other than the zero $F$-terms themselves); 
then different $X$ and $Y$ eigenvalues give different vacuum solutions. In general, a $d>1$ dimensional representation is not reducible if: $[X,Y]\neq 0$, the eigenvectors of $X$ and $Y$ 
collectively span at least a $d$-dimensional space, and $X$ and $Y$ do not share an eigenvector corresponding to a zero eigenvalue.

Consider a general $W_{A,D,E}$ theory, deformed by a generic $\Delta W$.  Let $i$ run over the vacuum solutions, and $d_i$ be their dimension.  There are always precisely $r_G\equiv \text{rank}(G)$ different $d_i=1$ dimensional (diagonalized) vacuum solutions for $X$ and $Y$, as with the original, $N_c=1$ Landau-Ginzburg theories \WArnold.  
  For the $D$ and $E$ cases, with $N_c>1$, there are $d_i>1$ dimensional vacuum solutions.  In all cases, the full $N_c\times N_c$ matrix expectation values of $X$ and $Y$ decompose into blocks, with $n_i$ copies of the $i$'th representation, such that
	\eqn{N_c=\sum _i n_i d_i.}[Ncblocks]
The vacua are given by all such partitions of $N_c$ into the $n_i$, subject to quantum stability constraints (to be discussed).  The non-zero $X$ and $Y$ Higgs $U(N_c)$ or $SU(N_c)$, with the unbroken gauge group depending on the $n_i$.   

It turns out that, if there are $n$ copies of a $d$-dimensional vacuum, there will be an unbroken $U(n)_{D}\subset U(N_c)$, where  $U(n)_{D}$ can be regarded as coming from breaking a $U(dn)\subset U(N_c)$ as $U(dn)\to U(n)^d\to U(n)_{D}$.  The $U(n)^d$ factors each have $N_f$ flavors, so the diagonally embedded $U(n)_{D}$ has $dN_f$ flavors.  If both adjoints receive a mass from the superpotential $F$-terms, the low-energy $U(n)_{D}$ will then be SQCD with $dN_f$ flavors.  This factor then has a  dual gauge group $U(dN_f-n)_{D}$, with $dN_f$ flavors (with $SU(N_f)_{L,R}$ enhanced to $SU(dN_f)_{L,R}$ as an accidental symmetry in the IR limit).   By the dual analog of the electric Higgsing, this low-energy $U(dN_f-n)_{D}$ can be embedded in a $U(d^2N_f-dn)$ with $N_f$ flavors.  For example, consider the case of $n$ copies of a 2d vacuum, with $\langle X\rangle$ breaking $U(2n)\to U(n)\times U(n)$, and then $\langle Y\rangle$ in the bifundamental breaking to $U(n)_{D}$.  Duality maps this process as follows:
	\eqna{
	\begin{array}{ccccc}
	U(2n)&\to & U(n)\times U(n) &\to & U(n)_{D} \\
	\big\downarrow  & &  & & \big\downarrow \\
	U(4N_f-2n)&\to &U(2N_f-n)\times U(2N_f-n) &\to & U(2N_f-n)_{D} 
	\end{array}
	}[higgsing]
The low-energy theory for such a vacuum is denoted as $A_1^{2d}$ if all the adjoints are massive, where the $2d$ superscript indicates that it comes from a 2d representation, and thus has $2N_f$ (or more generally, $dN_f$) flavors.   Applying such considerations for all $d_i$ vacua in \Ncblocks\ suggests that the dual theory has $\alpha$ given by
\eqn{\alpha \overset{?}{=}\sum _i d_i^2.}[sumofsquares]
This relation indeed works for the $A_k$ and the $D_\text{odd}$ theories, but not for $D_\text{even}$ or $E_{6,7,8}$.

For $W_{D_{k+2}}$, the generic deformation has $k+2$ 1d vacuum solutions, and $\lfloor \frac{1}{2}(k-1)\rfloor$ 2d representations.  If there are $n_i$ copies of the $i$'th 1d solution, and $n^{2d}_j$ copies of the $j$'th 2d solution, then $U(N_c)$ is broken as in \DkHiggsing.  For odd $k$, \sumofsquares\ indeed gives $\alpha = 3k$.

\newsec{Example and review: $\widehat A$ and $A_k$ one-adjoint cases \rcite{DKi,DKAS,DKNSAS,KPS}} \label{sec:oneadjsec}

\subsec{$\widehat A\to A_k$ flow and $A_k$ duality}[aduality]

Consider $SU(N_c)$ SQCD with $N_f$ chiral superfields $Q(\tilde{Q})$ in the (anti)fundamental of the gauge group, and adjoint chiral superfields   $X$ and $Y$  with superpotential 
	\eqn{
	W_{A_k} =  \frac{t_k}{k+1} \text{Tr}  X^{k+1} + \frac{m_Y}{2} \text{Tr} Y^2.
	}[WAk]
The $Y$ field is massive and can be integrated out; this is the $\widehat O\to \widehat A$ RG flow in Fig. \ref{fig:RGflows}.  The $t_k$ coupling, if relevant, drives the $\widehat A\to A_k$ RG flow in Fig. \ref{fig:RGflows}; if irrelevant, the IR theory is instead an $\widehat A$ SCFT.    For $k=1$, $t_{k}=m_X$ is an $X$ mass term and is always relevant; then both $X$ and $Y$ can be integrated out and the IR $A_1$ theory is ordinary SQCD.  For $k=2$, $t_k$ is marginally relevant as long as the matter content is within the asymptotically free range, thanks to the gauge coupling. For $k>2$, the $t_k$ coupling is relevant only if $x>x_{k}^{\text{min}}$ \rcite{KPS}.  

The chiral ring of the $A_k$ theory truncates classically, and we may write the $k$ generators
	\eqn{\Theta_{j} = X^{j-1},\ \ \ j=1,...,k. }[]
There are then $kN_f^2$ meson operators \mesonsj, with $\alpha _{A_k}= k$, and baryonic operators \baryons.  

The $\widehat A$ theory ($t_k=0$) does not have a known dual description.   The  magnetic description of the $A_k$ SCFT \rcite{DKi,DKAS,DKNSAS} has gauge group $SU(\tilde {N_c})$ with 
$\tilde{N_c}=kN_f-N_c$, so $\hat{x}\equiv \tilde N_c/N_f=k-x$. The dual has $N_f$ (anti)fundamentals $q (\tilde{q})$, adjoints $\hat{X},\hat{Y}$, and $k$ gauge singlets $M_j$ transforming in the bifundamental of the $SU(N_f)\times SU(N_f)$, with superpotential 
	\eqn{
	W_{A_k}^{mag} =  \frac{\hat{t}_k}{k+1} \text{Tr}  \hat{X}^{k+1} + \frac{\hat{m_Y}}{2} \text{Tr} \hat{Y}^2 +  \frac{t_k}{\mu^2} \sum_{j=1}^k M_j \tilde{q} \hat{X}^{k-j} q.
	}[WAkmag]
We can rescale $X$ and $\hat X$ to set $t_k=\hat t_k=1$, and $\mu$ is a scale that appears in the scale matching of the electric and magnetic theories. The $kN_f^2$ mesonic gauge invariant operators \mesonsj of the electric theory map to elementary gauge-singlets $M_j$ in the dual.  The other gauge invariant, composite operators in the chiral ring of the electric theory---i.e. the generalized baryons \baryons, operators $\tr X^{j-1}$, and glueball-type operators composed from $W_\alpha$---all map directly to the corresponding composite gauge-invariant chiral operators in the magnetic dual theory.  Both theories have the same anomaly free global symmetries, $SU(N_f)\times SU(N_f) \times U(1)_B \times U(1)_R$, and the 't Hooft anomalies properly match \rcite{DKi,DKAS,DKNSAS}. 

The $\widehat A$ theories have a quantum moduli space of vacua, $W_{dyn}=0$, for all $N_f$ and $N_c$.  The $A_k$ theories, however, generate $W_{dyn}\neq 0$ if $N_f<kN_c$.  For example, SQCD ($W_{A_1}$) for $N_f<N_c$ has $W_{dyn}\neq 0$~\rcite{Affleck:1983mk}, giving a~$\widetilde QQ\to \infty$ runaway instability for massless flavors or $\tr (-1)^F=N_c$ gapped susy vacua for massive flavors. We are here interested in cases with massless flavors and $W_{dyn}(M_j)=0$, so we restrict to $kN_f> N_c$, i.e. $x< x_\text{stability}=k$; this is the vacuum stability bound \rcite{DKi,DKAS}. For $kN_f<N_c$, the quantum theory $A_k$ has a moduli space of vacua, where the $M_j$ mesons have expectation values. The classical constraints on this moduli space, e.g. $\text{rank}(M_k)\leq N_c$, are recovered in the magnetic dual description from its stability bound, $\hat{x} <k$, since the $M_k$ expectation value gives masses via \WAkmag\ to the dual quarks $q$, $\widetilde q$.

\subsec{$W_{A_k}+\Delta W$ deformations and $A_k\to A_{k'<k}$ RG flows}[adeformations]

The $A_k$ theories of different $k$ are connected by RG flows upon resolving the $A_k$ singularity \WAk by lower order $\Delta W$ deformations.  The generic deformation, for instance by a mass term $\Delta W=\frac{1}{2} m_X\tr X^2$, leads to an RG flow with $k$ vacuum solutions for $\langle X\rangle$, with $X$ massive in each, hence $k$ copies of SQCD in the IR---i.e. $A_k\to kA_1$.   

We now consider a partial resolution,  
by tuning the superpotential couplings such that some of the eigenvalues coincide. We first consider the $U(N_c)$ case, which is simpler because we don't have to worry about imposing the tracelessness of $X$.  Consider the deformation
\eqn{
W_{elec}= W_{A_k}+\Delta W, \qquad \Delta W=\sum_{i=k'}^{k-1} \frac{t_i}{i+1} \text{Tr} X^{i+1}.
}[WAkAk]
(The $t_{k-1}$ deformation is trivial in the chiral ring, and it can be shifted away by shifting $X$, at the expense of inducing lower order terms.  Such items affect the RG flow, so we keep $t_{k-1}$ non-zero here.)  The $F$-terms of \WAkAk\ have a discrete set of solutions for the eigenvalues of $X$, with one solution at $X=0$ and $(k-k')$ solutions at non-zero values of $\langle X\rangle$.  

The vacua are given by all possible partitions of $N_c$ into the possible vacuum eigenvalues; in such a vacuum, the electric gauge group is broken as 
\eqn{
U(N_c)_{A_k}\to U(n_0) _{A_{k'}}\times \prod_{i=1}^{k-k'} U(n_i)_{A_1}, \qquad N_c=n_0+\sum _{i=1}^{k-k'} n_i.
}[AkAknodes]
The subscripts denote the low-energy theory, obtained by expanding \WAkAk\ around the corresponding vacuum, $X=\langle X\rangle +\delta X$.  The vacua at $\langle X\rangle =0$ have the most relevant term in \WAkAk\ given by $W_{low}\sim \tr (\delta X)^{k'+1}=W_{A_{k'}}$. The vacua at $\langle X\rangle \neq 0$ have a mass term for the low-energy adjoint, $W_{low}\sim \tr (\delta X)^2=W_{A_1}$.  We write this breaking pattern as
\eqn{A_k\to A_{k'}+(k-k')A_1.}[AktoAprimes]
By further tuning the $t_i$ parameters in the deformation \WAkAk, we could cause some or all of the $(k-k')$ SQCD vacua to coincide, e.g leading to 
\eqn{A_k\to A_{k'}+A_{k-k'}: \qquad \hbox{i.e.}\qquad U(N_c)_{A_k}\to U(n_0)_{A_{k'}}\times U(N_c-n_0)_{A_{k-k'}}.}[AktotwoAkprime]
Quantum mechanically, the vacuum stability condition---needed to have $W_{dyn}=0$---requires each $U(n)_{A_k}$ vacuum in \AkAknodes\ to have $kN_f>n$ \rcite{DKi,DKAS,DKNSAS}.

In the magnetic dual, we deform by the dual analog of the perturbations in \WAkAk.   The vacuum solutions of the deformed electric and dual theories, $W_{elec}'(X)=0$ and $W_{mag}'(\hat X)=0$, thus appropriately match, so if the electric breaking pattern is as in \AktoAprimes\ or \AktotwoAkprime, it will have the corresponding pattern in the magnetic dual.  Each vacuum gauge group in the low-energy theories maps under duality as \rcite{DKi,DKAS,DKNSAS}
	\eqn{U(n)_{A_k} \leftrightarrow U(kN_f-n)_{\widetilde{A_k}}}[]
and the stability bound in the electric theory ensures that $kN_f-n>0$.  The theories on the UV and IR sides of \AkAknodes\ thus map in the dual as 
\eqn{U(kN_f - N_c) _{\widetilde{A_k}}\to U(k'N_f - n_0) _{\widetilde{A_{k'}}}\times  \prod_{i=1}^{k-k'} U(N_f - n_i) _{\widetilde{A_1}}.}[AkAk]
For the case 
in \AktotwoAkprime\ the map is
\eqn{U(kN_f-N_c)_{\widetilde{A_k}}\to U(k'N_f-n_0)_{\widetilde{A_{k'}}}\times U((k-k')N_f-N_c+n_0)_{\widetilde{A_{k-k'}}}.}

The two sides of the RG flow arrow in \AkAk\ properly fit together as a dual description of the flow associated with the $\Delta W$ deformation, since $(k'N_f - n_0) + \sum_{i=1}^{k-k'} (N_f - n_i) = kN_f - \sum_{i=0}^{k-k'} n_i = kN_f-N_c$. This demonstrates that the value $\alpha _{A_k}=k$ (see Section \matrix) ties in with the fact that the $A_k$ deformation breaking patterns (e.g. as in 
\AktoAprimes) have matching sum on the two sides. This matching gives a check on the duality \rcite{DKNSAS} ---a perspective which we utilize throughout the present work. 

As an aside, we note that the $a$-theorem \atheorem applies for any choice of the IR vacuum; i.e. for any fixed choice of how to distribute the $N_c$ eigenvalues of $X$ among solutions to $W'(X)=0$ (subject to the stability bounds).   Regarding $a$ as counting a suitably defined ``number of degrees of freedom" of the QFT, one might wonder if a hypothetical stronger statement holds: if $a_{UV}$ is also larger than the sum $\sum _{i}a_{IR,i}$ over all IR vacua.  These examples show that the hypothetical stronger statement is false. There are so many vacua from the many partitions of $N_c$ that it is straightforward to explicitly verify that $\sum _{i}a_{IR,i}$ can be larger than $a_{UV}$.

\subsec{Comments on $SU(N_c)$ vs $U(N_c)$ RG flows}
\label{sec:su}

It is standard that the local\foot{Of course the global dynamics and observables distinguish the 
different center of $U(N_c)$ vs $SU(N_c)$.} dynamics of 4d $U(N_c)$ and $SU(N_c)$ are the same: the overall $U(1)$ factor in $U(N_c)$ is IR-free anyway in 4d (although that is not the case in 3d and lower).  The original dualities of \rcite{Seiberg:1994pq,DKi,DKAS,DKNSAS,JB} etc. were written in terms of $SU(N_c)$, with $U(1)_B$ as a global symmetry.  Since $U(1)_B$ is anomaly free, one can gauge it on both sides of the duality, leading to $U(N_c)\to U(\alpha N_f-N_c)$ dualities.  For the theories with adjoint matter, the $U(N_c)$ version of the theories are simpler, in that we do not need to impose the tracelessness of the adjoints. The adjoints $X$ of the $SU(N_c)$ vs $U(N_c)$ theories are related by $X_{U(N_c)}=X_{SU(N_c)}+X_0{\bf 1}_{N_c}$, where $\tr X_{SU(N_c)}=0$ and $X_0$ is an $SU(N_c)$ singlet.  In the purely $SU(N_c)$ theory, it is standard to eliminate $X_0$ by including a Lagrange multiplier $\lambda _x$: $W_{A_k}=\tr X^{k+1}/(k+1)-\lambda _x\tr X$. Then $\lambda_x$ pairs up with $X_0$, giving it a mass, and the vacua have $X_0=0$.  The $W_X=0$ chiral ring relation here gives $X^k=\lambda _x{\bf 1}_{N_c}$.

Upon deforming $W_{A,D,E}\to W_{A,D,E}+\Delta W$, the $\tr X_{SU(N_c)}=\tr Y_{SU(N_c)}=0$ constraints complicate the $SU(N_c)$ theories compared with $U(N_c)$.  This is particularly the case if we are interested in $\Delta W$ flows as in Fig. \ref{fig:RGflows} which have some $X$ and $Y$ dynamics remaining in the IR, rather than flowing all the way down to just decoupled copies of SQCD.  We can enforce $\tr X_{SU(N_c)}=\tr Y_{SU(N_c)}=0$ via Lagrange multipliers, which shifts the eigenvalues of $X$ and $Y$ along the flow away from the preferred $U(N_c)$ origin at $X=Y=0$. Such a shift will induce the more general, relevant $\Delta W$ deformations which were tuned to zero for the $U(N_c)$ case, unless the reintroduced $\Delta W$ terms are subtracted off by a tuned choice of coefficients in the initial $\Delta W$.  We will see that there are  subtleties---especially for the $D$ and $E$ cases---from the $d>1$ dimensional vacuum representations.  

Consider for example the flow $A_3\to A_2+A_1$. For $U(N_c)$, we get the enhanced $A_2$ in the IR (vs the generic $3A_1$) by taking $k'=2$ in \WAkAk:
	\eqna{
	W = \frac{1}{4} \text{Tr} X^4 + \frac{t_2}{3}  \text{Tr} X^3 + \frac{1}{2} \text{Tr} Y^2.
	}[aa]
For the $SU(N_c)$ version of this flow, we add  the Lagrange multiplier $\lambda _x$ to eliminate $X_0$, shifting the $X$ eigenvalues.  But simply doing this shift in \aa\ would induce the $\tr X^2$ term, giving instead $A_3\to 3A_1$.  To get $A_3\to A_2+A_1$, we must add to \aa\ the remaining $t_{m<2}$ terms in \WAkAk,
	\eqna{
	W=\frac{1}{4} \text{Tr} X^4+ \frac{t_2}{3}  \text{Tr} X^3+ \frac{1}{2} \text{Tr} Y^2  + \frac{t_1}{2} \text{Tr} X^2 - \lambda_x \text{Tr} X-\lambda _y\text{Tr}Y,
	}
with $t_1$ tuned in terms of the multiplicities $n_0,n_1$ of eigenvalues in the $A_2$ and $A_1$ solutions.  For fixed $t_1$, vacua with other partitions $N_c=n_0'+n_1'$ will instead have $3A_1$ in the IR. 

It is not immediately apparent if this procedure works in the $D$ and $E$ cases to shift higher-dimensional representations in just the right way to be able to map any $U(N_c)$ deformation into a corresponding $SU(N_c)$ one. The chiral ring algebra that determines how one labels the higher-dimensional vacua is sensitive to additional deformation terms in both $X$ and $Y$, with $[X,Y]\neq 0$.  While such a shift maps between the 1d $U(N_c)$ and $SU(N_c)$ solutions, the higher-dimensional solutions can differ; indeed, we will see examples of this later on. Additional subtleties arise when there are multiple ways to perform the shift between the 1d solutions of $SU(N_c)$ and $U(N_c)$.  We find cases in the $D$- and $E$-series where different deformation shifts agree for the 1d solutions but result in different Casimirs along the flow, thus affecting the labeling of higher-dimensional vacua. We will explore these issues with examples in Sections \ref{sec:Dshift} and \ref{sec:Eflows}.

\subsec{$SU(N_c)$ flat direction deformations}[FDAk]
\label{sec:FDAk}

The ADE SCFTs, for $SU(N_c)$ gauge group and special values of $N_c$, have flat directions that are not present for $U(N_c)$. These are discussed for the $A_k$ case in \rcite{DKNSAS}. Adding a Lagrange multiplier term $\lambda_x \text{Tr} X$ to \WAk, there is a flat direction of supersymmetric vacua when $N_c=km$ for integer $m$, labeled by arbitrary complex $\lambda_x$:
	\eqn{
	\langle X \rangle = \lambda_x^{1/k} \left( \begin{array}{cccc} \omega {\bf 1}_m &  &  &  \\  & \omega^2 {\bf 1}_m &   &  \\  &  & \ddots &  \\  &  &  & \omega^k {\bf 1}_m \end{array} \right),
	}
where  $\omega = e^{2\pi i/k}$ is a $k$'th root of unity and the off-diagonals are zero. This flat direction breaks $SU(N_c)\to SU(m)^k\times U(1)^{k-1}$. In each vacuum the adjoints are massive, so in the IR we end up with $k$ copies of SQCD. The magnetic $A_k$ theory has an analogous flat direction,  along which the low-energy theory  matches to that of the $k$ copies of SQCD via Seiberg duality:
	\eqna{
	\begin{array}{ccc}
	SU(km) & \overset{\lambda_x \neq 0}{\longrightarrow} & SU(m)^k\times U(1)^{k-1} \\ 
	  \big\downarrow\ \ \ \ \ \ \  & & \big\downarrow \\ 
	SU(k(N_f-m)) & \to & SU(N_f -m)^k \times U(1)^{k-1} 
	\end{array}}[akfdduality]
This gives yet another check that the $A_k$ duality has $\tilde{N_c} = \alpha N_f - N_c$, with $\alpha=k$.

\newsec{The $W_{D_{k+2}}$ fixed points and flows}\label{sec:D}

The $W_{D_{k+2}}$ SCFTs are the IR endpoints of the RG flow from the $\hat{D}$ SCFT, and correspond to the superpotential (with $Y$ normalized to set the coefficient of the first term to 1)
	\eqn{
	W_{D_{k+2}} = \text{Tr} X Y^2 + \frac{t_k}{k+1} \text{Tr} X^{k+1}.
	}[WDk]
Such theories were first studied in \rcite{JB}. The $\text{Tr}XY^2$ term in \WDk\ is always relevant and drives the RG flow $\widehat O\to \widehat D$, while the second term in \WDk\ gives the $\widehat D\to D_{k+2}$ RG flow.  For $k=1$, $W_{D_3}\cong W_{A_3}$, since then \WDk\ contains the (relevant) $X$-mass term $\text{Tr}X^2$, and integrating out $X$ yields $W_{low}\sim \text{Tr} Y^4\sim W_{A_3}$.  For $k=2$, the superpotential \WDk\ is cubic, and hence relevant as long as the gauge group is asymptotically free, i.e. $x>1$.  For $k>2$, 
the $\widehat D\to D_{k+2}$ flow associated with the coupling $t_k$ is relevant only if $x>x_{D_{k+2}}^{\text{min}}$, where $x_{D_{k+2}}^{\text{min}}$ was determined via $a$-maximization in \rcite{twoadj},  
	\eqn{
	x_{D_{k+2}}^{\text{min}}\ \ \bigg\{ \begin{array}{c l l} = & \frac{1}{3\sqrt{2}} \sqrt{10-34k+19k^2} & k<5 \\ < & \frac{9}{8} (k+1) & k\ \text{large} \end{array}.
	}[xmin]
For relevant $t_k$, we can normalize $X$ to set $t_k=1$ at the IR $D_{k+2}$ SCFT.  For $x<x_{D_{k+2}}^\text{min}$, $t_k\to 0$ in the IR and the theory stays at the $\widehat D$ SCFT.  We will here assume that $x>x_{D_{k+2}}^\text{min}$.

The F-terms of the undeformed $D_{k+2}$ superpotential \WDk\ are given by
	\eqna{
	Y^2 + t_k X^k = 0,\\
	\{X,Y\} = 0.
	}[DFterms]
For $k$ odd, it follows from \DFterms\ (as explained after \Dfterm) that the chiral ring classically truncates to the $3k$ generators \Dring. As in the $A_k$ case, there is a stability bound: we must require $x<x_{\rm stability}$ in order to avoid $W_{dyn}$, which would lead to a runaway potential for the generalized mesons. For $x<x_{\rm stability}$, there is instead a moduli space of supersymmetric vacua with $W_{dyn}=0$.  As we will review (at least for odd $k$) $x_{\rm stability}=3k$, which is related to the fact that the chiral ring has $3k$ elements.

\subsec{Proposed dualities for $W_{D_{k+2}}$ \rcite{JB}}[pd]

A dual description of the $D_{k+2}$ theories was proposed in \rcite{JB}, and many of the usual, non-trivial checks were verified---for instance matching of the global symmetries, 't Hooft anomaly matching, and mapping of the chiral ring operators. As reviewed in Section \matrix, the conjectured duals have gauge group $SU(\alpha_{D_{k+2}} N_f-N_c)$ with $\alpha _{D_{k+2}}=3k$ \Ddualalpha, and matter content consisting of $N_f$ (anti)fundamentals $q (\tilde{q})$, adjoints $\hat{X},\hat{Y}$, and $3k$ gauge singlet mesons $M_{jl}$ which map to the composite meson operators of the electric theory as
	 \eqn{
	M_{\ell j} = \tilde{Q} X^{\ell-1}Y^{j-1} Q, \quad \ell=1,...,k;\ \ j = 1,2,3.
	}[dmesons]
The dual theory has superpotential 
  	\eqn{
 	W_{D_{k+2}}^{mag}= \text{Tr} \hat{X}\hat{Y}^2 + \frac{1}{k+1} \text{Tr} \hat{X}^{k+1} + \frac{1}{\mu^4} \sum_{\ell=1}^k \sum_{j=1}^3 M_{\ell j} \tilde{q} \hat{X}^{k-\ell}\hat{Y}^{3-j}q.
 	}[ddualsup]
A detailed calculation, via $a$-maximization, is needed to determine the $\hat x_{min}$  \hatx values for the various non-cubic terms in \ddualsup\ to be relevant rather than irrelevant \rcite{twoadj}. 

The above dual, with $\alpha _{D_{k+2}}=3k$ mesonic operators \dmesons, requires the chiral ring truncation \Dring, which is  only evident from the classical $F$-terms for $k$ odd. It was conjectured in \rcite{JB} that quantum effects make the even $k$ theories similar to odd $k$, with a quantum truncation of the chiral ring, in order for the duality to hold for both even and odd $k$.  It is as-yet unknown if and how such a quantum truncation occurs for the even $k$ case, and thus the status of the duality remains  uncertain for even $k$.  The fact that e.g. the 't Hooft anomaly matching checks work irrespective of whether $k$ is even or odd can be viewed as evidence that the duality also applies for $D_{\text{even}}$,  or perhaps just  a  coincidence following merely from the fact that these checks are meaningful for odd $k$.

In addition to the usual checks of duality, the proposed chiral ring truncation and duality for $D_{\text{even}}$ were used in \rcite{JB} to predict a duality for an $SU(N_c)\times SU(N_c')$ quiver gauge theory with (anti)fundamentals and an adjoint for each node, and (anti)bifundamentals between. This latter duality was later re-derived, and confirmed, by considering deformations of the more solid, odd $k$ $D_{k+2}$ theories \rcite{Brodie:1996xm}.  
But it was also noted in \rcite{Brodie:1996xm} that the $D_{k+2}$ duality implies some other dualities that are clearly only applicable for $k$ odd, with fields appearing in the superpotentials with powers like $X^{(k+1)/2}$.  The fractional power for $k$-even suggests an incomplete description, which is missing some additional degrees of freedom.  The status of the $D_\text{even}$ duality thus remained (and it still remains) inconclusive.

A powerful, more recent check of dualities is to verify that the superconformal indices of the electric and magnetic theories match; see e.g. \rcite{Dolan:2008qi,Spiridonov:2009za}. In \rcite{Kutasov:2014wwa}, the superconformal indices for the electric and magnetic dual $D_{k+2}$ theories are verified to indeed match in the Veneziano limit for both even and odd $k$.  The matching beyond the Veneziano limit provides a physical basis for a conjectural mathematical identity.   It was moreover noted in \rcite{Kutasov:2014wwa} that the conjectural quantum truncation of the $k$ even chiral ring should be verifiable via the the index, by expanding it to the appropriate order in the fugacities and checking if the contributions from operators that are eliminated by the quantum constraints are indeed cancelled by those of other operators.  It was noted, however, that this check is complicated by the fact that there are many possible contributing operators, so it was not yet completed.

One of the original arguments for the $D_{\text{even}}$ quantum truncation is based on the fact that one can RG flow from  $D_{\text{odd}}\to D_{\text{even}}$ via 
appropriate $\Delta W$ deformation, e.g. $D_{k+2}\to D_{k+1}+A_1$.  
Another, similar argument \rcite{twoadj} uses the connection between the stability bound and the chiral ring truncation. The duality suggests that the original electric theory has an instability, e.g. via $W_{dyn}\neq 0$ leading to a runaway vacuum instability, when $3kN_f-N_c<0$, i.e. for $x>3k$, and we expect RG flows to reduce the stability bound in the IR.  Flowing, for instance, from $D_{k+2}\to D_{k+1}+A_1$ for $k$ odd, the UV $D_{k+2}$ theory has a truncated chiral ring and stability bound, which suggests that the IR (even) $D_{k+1}$ theory should also have a stability bound, and hence chiral ring truncation. We will analyze such RG flows in detail here, and show that there are subtleties.  

In summary, the evidence that the duality holds for $D_{\text{odd}}$ is compelling, while the evidence for $D_{\text{even}}$ is mixed, with aspects that are not understood.  Our analysis here fails to find evidence for the quantum truncation of the chiral ring for $D_{\text{even}}$, and instead points out additional hurdles for the conjectured duality.

\subsec{Matrix-related flat directions at the origin}

\subsubsec{A 2d line of flat directions for $D_{\text{even}}$}[dline]

We consider the moduli space of vacuum solutions of \DFterms,  and the $D$-term constraints \Dterms, taking $X, Y\neq 0$ with $Q=\tilde Q=0$.  The 1d versions of these equations, where we replace the matrices with 1d eigenvalue variables $X\to x$, $Y\to y$, are only solved at $x=y=0$, corresponding to the $D_{k+2}$ singularity at the origin of the moduli space for the undeformed $W_{D_{k+2}}$ theory.   Now consider $d>1$ dimensional representations of the solutions of \DFterms\ and \Dterms.  The second $F$-term shows that $[X^2,Y]=0$, so $X^2$ is a Casimir.  Likewise, it follows from the first $F$-term in \DFterms\ that $[Y^2,X]=0$, so $Y^2$ is also a Casimir; the representation must have $X^2=x^2{\bf 1}_d$, and $Y^2=y^2{\bf 1}_d$.  For $D_\text{odd}$, the first $F$-term in \DFterms\ would then imply that $X$ is also a Casimir, so there can not be a non-trivial $d>1$ dimensional representation.  For $D_\text{even}$, on the other hand, the $F$-terms, $D$-terms, and Casimir conditions are solved by the $2$-dimensional solutions
	\eqna{
	k\ \text{even}:\quad X = x\sigma_3,\quad Y=y\sigma_1 \\
	y^2 + t_k x^k = 0.
	}[FDline]
This gives a moduli space of supersymmetric vacua, passing through the origin.  Modding out by gauge transformations, which take $x\to -x$ and $y\to -y$, the moduli space can be labeled by 
$x^2$ and $y^2$ satisfying \FDline, which allows for an additional ${\bf Z}_{k/2}$ phase for $x^2$.  Since $X$ and $Y$ in \FDline\ are traceless, this flat direction is present for either $SU(N_c)$ or $U(N_c)$.

More generally, $D_\text{even}$ has vacua with multiple copies of the 2d vacuum solution \FDline,
with the remaining eigenvalues of $X$ and $Y$ at the origin.   There can be $\lfloor N_c/2\rfloor$ copies of the 2d representation, giving a moduli space of supersymmetric vacua labelled by  $x_i^2$ and $y_i^2$ satisfying \FDline, for $i=1,\dots, \lfloor N_c/2\rfloor$.  The $SU(N_f)_L\times SU(N_f)_R$ global symmetries are unbroken along this subspace, so it can be distinguished from the mesonic or baryonic directions where the $Q_f$ or $\tilde Q_{\tilde f}$ have expectation value.  The classically unbroken gauge symmetry is enhanced when various $x_i$ are either zero or equal to each other.  Consider, for example, $N_c=2n$, with all $n$ of the $x_i$ non-zero and equal.  In this direction of the moduli space, by a  similarity transformation, we have
	\eqna{
	\langle X \rangle &= x {\bf 1}_n\otimes \sigma _3\overset{\ B\langle X\rangle B^{-1}}{ \longrightarrow} x \sigma _3\otimes {\bf 1}_n\\
	\langle Y \rangle  &=  y{\bf 1}_n\otimes \sigma _1 \overset{\ B\langle Y\rangle B^{-1}}{ \longrightarrow}  y \sigma _1\otimes{\bf 1}_n.}[simtransf]
Consider the Higgsing in stages: first, $\langle X \rangle$ breaks $U(2n)\to U(n)\times U(n)$, and then $\langle Y\rangle$ breaks $U(n)\times U(n)\to U(n)_D$, the diagonally embedded subgroup (for simplicity, we write the gauge groups as $U(m)$, and corresponding expressions apply if we work in terms of $SU(m)$ groups).   This breaking pattern leaves five uneaten $U(n)_D$ adjoints from $X$ and $Y$, four of which get a mass from the $W_{D_{k+2}}$ superpotential \WDk.  The low-energy $U(n)_D$ along this moduli space has a massless adjoint matter field and $W_{low}=0$; i.e. it is a $U(n)_D$ $\widehat A$ theory.  Giving general expectation values to the  adjoint matter field of the low-energy $\widehat A$ theory corresponds to unequal expectation values of the $x_i$ in the  $n$ copies of the 2d vacuum  \FDline,  leading to the more generic breaking pattern $U(2n)\to U(n)_D\to U(1)^n$.   Note also that the low-energy $U(n)_D$ $\widehat A$ theory, along the moduli space \simtransf\ has $N_f^{low}=2N_f$ flavors, since the fundamentals decompose as ${\bf 2n}\to {\bf (n,1)+(1,n)}\to 2\cdot {\bf n}$; the enhanced flavor symmetry arises as an accidental symmetry.  In summary, there is a (classical) flat direction
\eqn{D_{k+2=\text{even}} \to \widehat A, \qquad\hbox{with} \quad U(N_c)\to U(\lfloor N_c/2\rfloor)_D \quad\hbox{and}\quad  N_f^{low}=2N_f,}[DevenAhat]
so $x^{low}=N_c^{low}/N_f^{low}=(N_c/2)/(2N_f)=x/4$.

We have not found a mechanism for this classical moduli space to be lifted by a dynamical superpotential or removed by quantum effects.   The low-energy $U(\lfloor N_c/2\rfloor)_{\widehat A}$ theory with $2N_f$ flavors clearly has $W_{dyn}=0$, and unmodified  quantum moduli space. The original theory can have additional effects e.g. from instantons in the broken part of the group (see \rcite{Intriligator:1995id,Csaki:1998vv} for discussion and examples), from the last step of the breaking $U(N_c)\to U(\lfloor N_c/2\rfloor)^2\to U(\lfloor N_c/2\rfloor)_D$ in \DevenAhat. Indeed, for $x$ above the stability bound, there can be a $W_{dyn}$ which leads to runaway expectation values for the mesonic operators.  But holomorphy, the $U(1)_R$ symmetry, and the condition that $W_{dyn}$ must lead to a potential that, by asymptotic freedom, goes to zero far from the origin of the moduli space, precludes any $W_{dyn}$ that only lifts the 2d flat directions \FDline without generating a runaway $W_{dyn}$ for the mesonic operators. As usual, the low-energy theory along the flat direction is less asymptotically free than the theory at the origin, and the theory is more weakly coupled for vacua farther from the origin on the moduli space.    The original $D_\text{even}$ theory at the origin is asymptotically free for $N_f<N_c$, while the low-energy $U(\lfloor N_c/2\rfloor)_{\widehat A}$ theory far along the flat direction is IR-free if $N_f>(N_c/2)$, i.e. if $x<2$.  In that case, the IR spectrum consists of the IR-free $U(\lfloor N_c/2\rfloor)_{\widehat A}$ gauge fields and matter. 

We now consider if this $D_\text{even}$ flat direction is compatible with the conjectural, dual $U(3kN_f-N_c)_{D_\text{even}}$ theory.  That theory has an analogous moduli space of vacua  where the dual adjoints $\hat{X},\hat{Y}$ satisfy $F$-term equations analogous to \DFterms, with copies of the 2d representation \FDline.  Chiral ring elements like  $\text{Tr} X^n$ should indeed map to similar elements in the dual, e.g. $\text{Tr} X^n\leftrightarrow \text{Tr}\widehat X^n$.  The moduli space of eigenvalues of the 2d representation is  $\frac{1}{2}(3kN_f-N_c)$-dimensional, along which the gauge group is broken to $U( \lfloor \frac{1}{2}(3kN_f-N_c)\rfloor)_{\widehat A}$.    The dimensions of the two moduli spaces differ, which is a contradiction with the conjectural dual unless some quantum effect eliminates the difference (as indeed happens with the mesonic directions of the moduli space, where the classical constraints on the rank of the meson matrices arise from quantum effects in the dual).
In addition to the moduli spaces differing, the low-energy theories on the flat directions of the two conjectured duals, i.e. $U(\lfloor N_c/2 \rfloor )_{\widehat A}$ and $U(\lfloor 3kN_f/2-N_c/2 \rfloor )$, are not in any clear way dual to each other; there is no known dual for the $W_{\widehat A}$ SCFTs.

As in the electric theory, we do not yet see a mechanism for quantum effects to modify the classical dimensions of these moduli spaces.  Note that the low-energy $U(\lfloor3kN_f/2-N_c/2\rfloor)_{\widehat A}$ theory is IR-free if $x>3k-2$, which is non-overlapping with the range $x<2$ where the corresponding electric theory is IR-free; this at least avoids an immediate, sharp contradiction with the duality, since two theories cannot have a different IR-free spectrum in the same region of the moduli space.   As a concrete example, consider the case $k=2$, i.e. $W_{D_4}$, and take $N_c$ even.  The electric $W_{D_4}$ superpotential \WDk\ is relevant as long as the gauge group is asymptotically free, for $x>1$.  The stability bound suggested by the conjectural $U(6N_f-N_c)$ dual is $x<6$.  The electric theory has the flat direction to the low-energy $U(N_c/2)_{\widehat A}$ theory, which is IR-free if $x<2$.  The dual theory has a flat direction to a low-energy $U(3N_f-N_c/2)_{\widehat A}$ theory with $2N_f$ flavors, which is IR-free if $x>4$.

This $D_\text{even}$ flat direction is related to the fact that the chiral ring of the $D_{\text{even}}$ theory does not classically truncate; one can think of it as coming from the massless degrees of freedom present in the non-truncated ring. Its existence provides us with a new way to rephrase the puzzle of how the truncation occurs: does some quantum effect lift this flat direction? If not, the flat direction seems inconsistent with duality.

\subsubsec{A puzzle for the $W_{D_\text{even}}$ flat directions \FDline: apparent $a$-theorem violations}[athm]

The supersymmetric flat direction discussed in the previous subsection has another puzzle, independent of the conjectured duality: it leads to naive violations of the $a$-theorem \atheorem for sufficiently large $x$.  The exact $a_{SCFT}$ is evaluated by using the relation \aRsusy\ between $a$ and the 't Hooft anomalies for the superconformal $U(1)_R$ symmetry, along with $a$-maximization (when needed) and accounting for all accidental symmetries.  The values of $a_{SCFT}$ for the $W_{D_{k+2}}$ theories were analyzed in \rcite{twoadj}, following the $W_{A_k}$ analysis in \rcite{KPS} with regard to the crucial role of including the effect of accidental symmetries in $a$-maximization.  One type of accidental symmetry, when gauge invariant chiral operators hit the unitarity bound and decouple, is readily apparent in the electric theory. 
Dualities reveal other types of accidental symmetries, e.g. those where the analog of the $t_k$ coupling in \WDk\ for the magnetic dual is irrelevant, or where the magnetic gauge coupling is irrelevant (the free-magnetic phase); such accidental symmetries are---as far as we know---unseen without knowing the dual.  

We consider $\Delta a$ for the RG flow associated with the flat direction in \DevenAhat. We compute $a_{UV}(x)$ corresponding to the $D_{k+2}$ theory with gauge group $SU(N_c)$ and $N_f$ flavors as in \rcite{twoadj}, and $a_{IR}(x)$ corresponding to an $\widehat{A}$ theory with gauge group $SU(N_c/2)$ and $2N_f$ flavors as in \rcite{KPS}, including as there the effects of all mesons hitting the unitarity bound and becoming IR-free.  We plot the results for the cases $k=2$ and $k=4$, working in the Veneziano limit of large $N_c$ and $N_f$, with $x$ fixed. ($U(N_c)$ vs $SU(N_c)$ is a subleading difference in this limit.)

 \begin{figure}[h]
\centering
	\begin{subfigure}[h]{0.495\textwidth}
		\includegraphics[width=\textwidth]{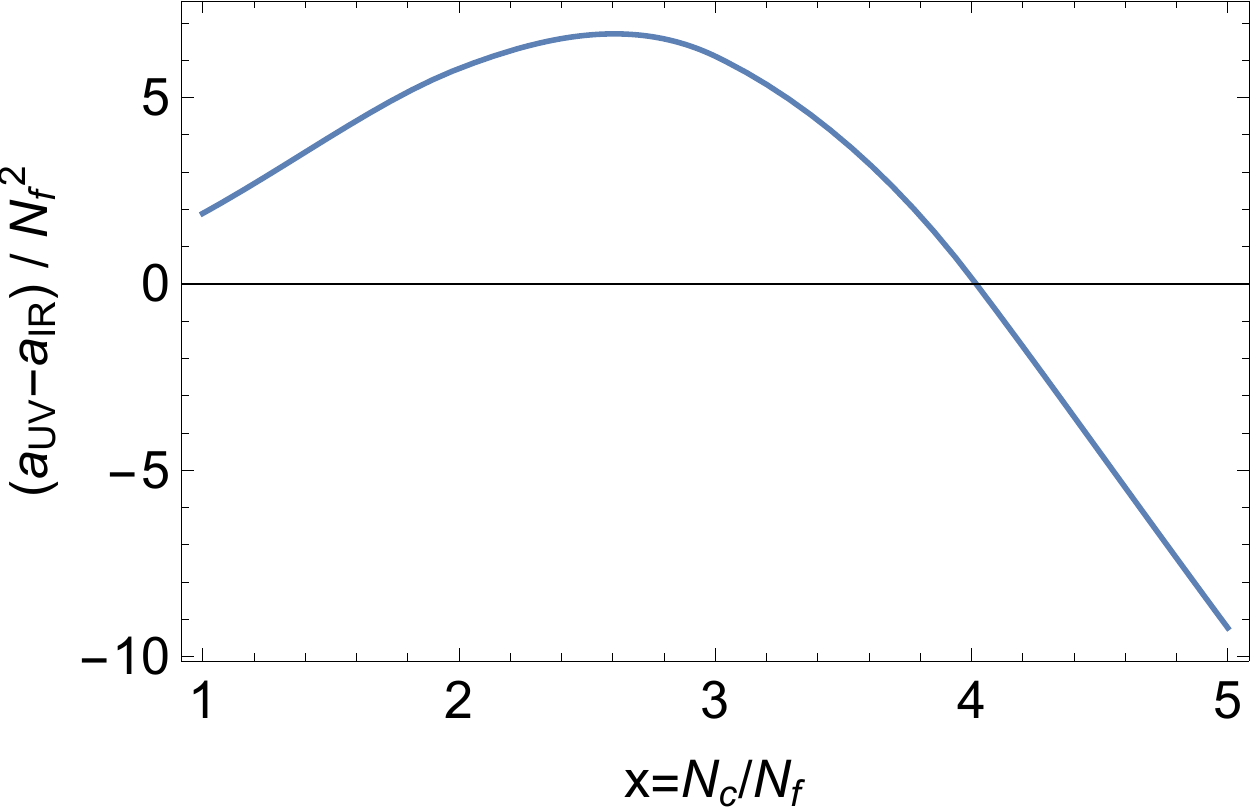}
		\caption{$k=2$ \label{fig:d4}}
	\end{subfigure}
	\begin{subfigure}[h]{0.495\textwidth}
		\includegraphics[width=\textwidth]{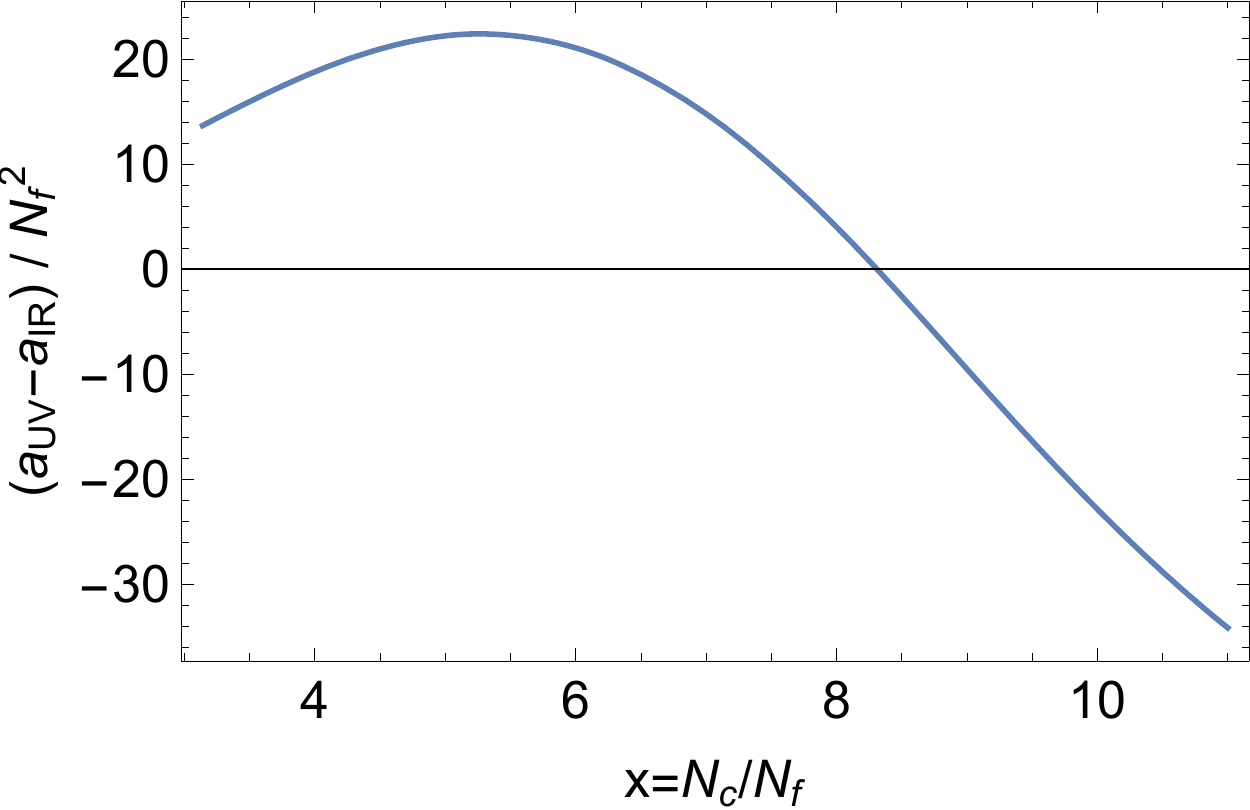}
		\caption{$k=4$\label{fig:d6}}
	\end{subfigure} 
\caption{$(a_{UV}-a_{IR})/N_f^2$ for (a) $W_{D_4}\to \widehat A$ and (b) $W_{D_6}\to\widehat A$, plotted for $x$ in the respective conformal windows. The $W_{D_4}$ theory is IR-free for $x\leq 1$. The $W_{D_6}$ theory requires $x>3.14$ for the $\text{Tr}X^5$ term in $W_{D_6}$ to be relevant, while the corresponding term in the Brodie dual is relevant if  $x<8.93$. The $\widehat{A}$ theory is asymptotically free in both plotted domains.}\label{fig:atheoremviolate}
\end{figure}

As we can see in Figures  \ref{fig:d4} and \ref{fig:d6}, both the $k=2$ and the $k=4$ flat direction RG flows seem to violate the $a$-theorem for sufficiently large $x$.  For the $W_{D_4}$ case, the conformal window where both the electric and magnetic theories are asymptotically free is $1<x<5$, and the cubic $t_k$ term in 
\WDk, or its magnetic analog, is relevant in this entire $x$ range. As seen in Fig. \ref{fig:d4}, the $a$-theorem is seemingly violated for $x\geq 4$, within the conformal window.  For $W_{D_6}$, the situation is plotted in Fig. \ref{fig:d6}: the flat direction seemingly violates the $a$-theorem for $x\geq 8.31$.  This is within the expected $W_{D_6}$ conformal window (i.e. below $3k-\hat{x}_\text{min}\approx 8.93$ beyond which duality suggests that the theory is instead in the $\widehat D_{mag}$ phase, and below $x=11$, where duality suggests  the IR-free magnetic phase). 

Of course, we do not believe that there will be violations of the $a$-theorem, so the puzzle of these apparent violations must somehow be resolved.  We also note that the apparent violations first occur for $x$ still below the values where mesons involving $Y^3$ would first hit the unitarity bound (this occurs first at $x=5$ for $k=2$, and at $x=9.33$ for $k=4$).  Thus, the calculation of $a_{UV}$ is not affected by the issue of whether or not such mesons should be included---we've removed them in the plots above, which would be correct if Brodie duality is correct for $D_{\text{even}}$ and the quantum truncation indeed occurs. 

We see two possible resolutions to the puzzle of the apparent $a$-theorem violations.   1) These classical flat directions are somehow lifted by quantum effects, in a way that we do not yet understand. 
 2) Some additional degrees of freedom make the calculation of $a$ wrong, e.g. giving a larger value for $a_{UV}$ for the $W_{D_\text{even}}$ theory.  We do not yet know the resolution. 
 
Option 1) could also resolve the conflict with Brodie-duality, discussed in the previous subsection.  As we discussed there, asymptotic freedom, along with holomorphy and the R-symmetry, suggests that $W_{exact}=0$, but perhaps another mechanism could remove the flat directions---at least for $x$ large enough to be in the problematic range.  The existence of the classical flat direction fits with the classically untruncated chiral ring, and it sharpens the issue of if, and how, the  chiral ring for the $D_{\text{even}}$ theory is quantumly truncated.

\subsubsec{Additional evidence that the $W_{D_\text{even}}\to \widehat A$ flat directions aren't lifted}[qflat]

We here present additional arguments against any quantum barrier to the $W_{D_\text{even}}\to W_{\widehat A}$ flat directions.  The idea is to explore more of the full moduli space of supersymmetric vacua, going along $Q$-flat directions, until the low-energy theory is IR-free.   

Consider an even $D_{k+2}$ theory at the origin, with $N_f<N_c$ such that the theory is asymptotically free.  Going along a $Q$-flat direction by giving a vev to a flavor, $\langle Q_f \rangle = (v,0,...,0) = \langle \tilde{Q}_f \rangle$, gives a low-energy theory that is less asymptotically free. The gauge group is Higgsed $SU(N_c)\to SU(N_c-1)$, under which the adjoints decompose $X\to \check X+F_x+\tilde F_x+s_x$ for $\check X$ an adjoint and $s_x$ a singlet (and likewise for $Y$). Then, the number of light flavors in the low-energy theory is $N_f-1+2=N_f+1$, where the $-1$ is for the eaten flavor and the $+2$ is from additional light flavors, $F_{x,y}$. Expanding the superpotential under this decomposition gives, for instance for $W_{D_4}$, an IR superpotential of the form
	\eqna{
	W_{D_4} &= \text{Tr} \bigg( \frac{1}{3} \check{X}^3 + \check{X} \check{Y}^2 + \check{X} \tilde{F}_x F_x + \check{X} \tilde{F}_y F_y + \check{Y} \tilde{F}_y F_x + \check{Y} \tilde{F}_x F_y +s_x F_x \tilde{F}_x + s_x F_y \tilde{F}_y\\
	&\ \ \ \ \ \ \ \  + s_y F_x \tilde{F}_y + s_y F_y \tilde{F}_x + s_x^3 + s_x s_y^2\bigg).
	}[Dexpand]
Along the above flat direction, the 1-loop beta function coefficient changes by $b_1=N_c-N_f\to (N_c-1)-(N_f+1)=b_1-2$ so, as usual, the low-energy theory is less asymptotically free.  We iterate this procedure, giving expectation values to $n$ flavors of $Q$ and $\tilde Q$, and thus reducing $N_c\to N_c-n$, with $N_f\to N_f+n$ and $b_1\to b_1-2n$, until the low-energy theory is no longer asymptotically free, i.e. $n>(N_c-N_f)/2$.  Then $X$ decomposes as
	\eqna{
	X &\longrightarrow   \left(\begin{array}{ccc|c} s_x^1 &  & &  F_x^1   \\ & \ddots & &  \vdots  \\ & & s_x^n &   F_x^n   \\ \hline  \tilde{F}_x^1 & ... & \tilde{F}_x^n &  \check{X}    \end{array}\right)}
with $\check{X}$ adjoints of an unbroken $SU(N_c-n)$, and similarly for $Y$.  

At this point, we can take $\check{X}$ and $\check{Y}$ in the low-energy $SU(N_c-n)$ theory to have an expectation value with $m\leq (N_c-n)/2$ copies of the 2d vev \FDline, resulting in the $\widehat{A}$ flat direction where $SU(N_c-n)\to SU(m)_D\times SU(N_c-n-2m)$.  By choice of $n$, the intermediate $SU(N_c-n)$ theory is already IR-free, and so the  $\check{X}$ and $\check{Y}$ expectation values make the low-energy theory even more weakly coupled; thus, the terms in $W_{low}$ (e.g. in \Dexpand) involving the singlets and fundamentals are irrelevant and can be ignored. The number of flavors of the low-energy $SU(m)_D$ theory is $2(N_f+n-r)$, where  $N_f+n$ flavors came from the $n$ iterations of $Q$-Higgsing, $r\leq n$ is the number of the $F_{x,y}$ flavors that receive a mass from $\langle \check{X}\rangle,\langle \check{Y}\rangle$ in the superpotential, and the 2 comes from Higgsing $SU(2m)\to SU(m)\times SU(m)\to SU(m)_D$. By taking $m$ sufficiently small and $n$ sufficiently large, the low-energy $SU(m)_D$  $W_{\widehat{A}}$ theory will have a 1-loop beta function of non-asymptotically free sign, so the theory will be IR-free and thus weakly coupled.  Because every interaction is IR-free in this region of the moduli space, quantum effects from the intermediate or low-energy theory cannot lift or remove the $W_{D_\text{even}}\to W_{\widehat A}$ flat direction.  As remarked earlier, any possible effects from the Higgsed, original gauge theory at the origin (e.g. instantons in the broken part of the group) must moreover slope to zero for vacua farther from the origin on the classical moduli space \FDline.  

\begin{figure}[t]
\centering
	\includegraphics[width=0.4\textwidth]{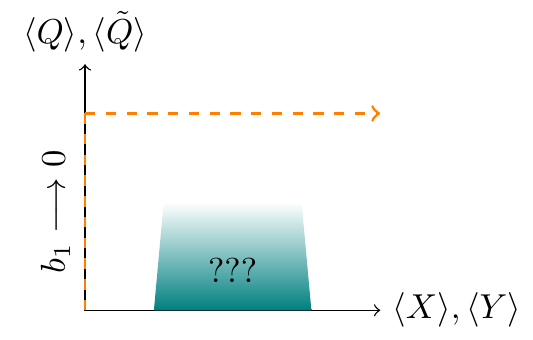}
	\caption{Using the $Q$-flat directions to bypass the strong coupling regime.}
\label{fig:qdrama}
\end{figure} 

In sum, as illustrated in Fig. \ref{fig:qdrama}, we use the $Q$-flat directions to bypass any hypothetical quantum barrier to the flat directions \FDline by going to a region of moduli space where the theory is IR-free.  This suggests that the $W_{D_\text{even}}\to W_{\widehat A}$ moduli space is indeed present in the full, quantum theory.  As discussed in the previous subsection, there would then have to be some missing contribution to $a$ for the $D_{\text{even}}$ theory to avoid the apparent $a$-theorem violation along this moduli space for sufficiently large $x$.

\subsec{$SU(N_c)$-specific (as opposed to $U(N_c))$ flat directions}[dsun]

\begin{figure}[h!]
\centering
	\begin{subfigure}[h]{0.69\textwidth}
		\includegraphics[width=\textwidth]{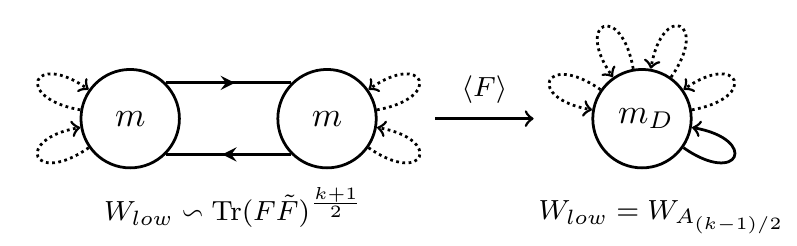}
		\caption{$SU(m)\times SU(m)$ sector.}
		\label{}
	\end{subfigure}\\
	\begin{subfigure}[h]{0.39\textwidth}
		\includegraphics[width=\textwidth]{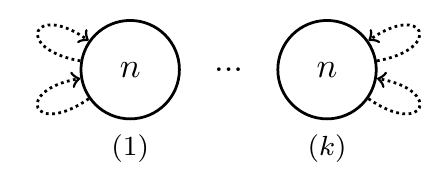}
		\caption{$SU(n)^k$ sector.}
		\label{}
	\end{subfigure}
\caption{Flat directions for $D_\text{odd}$, $N_c=2m+kn$, integrating out massive fields (denoted by dotted lines) and fields eaten by the Higgs mechanism (not shown).}
\label{fig:FLATodd}
\end{figure}

For $SU(N_c)$, one includes Lagrange multipliers $\lambda _x,\lambda _y$ to impose $\text{Tr}X=\text{Tr}Y=0$:
\eqn{
W_{D_k+2}= \text{Tr} XY^2 + \frac{t_k}{k+1} \text{Tr} X^{k+1} -\lambda_x \text{Tr} X - \lambda_y \text{Tr} Y.
}[dflatsup]
For $D_\text{odd}$, and  $N_c=2m+kn$ for $m,n$ integers, there is a flat direction labeled by $\lambda_x$ \rcite{JB} 
	\eqna{
	\langle X \rangle = \left(\frac{\lambda_x}{t_k}\right)^{\frac{1}{k}} &\left(\begin{array}{ccccc} {\bf 0}_m & & & & \\ & {\bf 0}_m & & & \\ 
  & &  {\bf 1}_n & & \\  & & & \ddots & \\  & & & & \omega^{k-1}  {\bf 1}_n\end{array}\right),\\
  \langle Y \rangle = \left(\lambda_x \right)^{\frac{1}{2}} &\left(\begin{array}{ccccc} {\bf 1}_m & & & &  \\ & -{\bf 1}_m & & &  \\ & & {\bf 0}_n & &  \\  & & & \ddots & \\  & & & & {\bf 0}_n \end{array}\right),
	}[Brodievev]
where $\omega = e^{2\pi i/k}$. The gauge group is Higgsed as $SU(2m+kn)\to SU(m)^2\times SU(n)^k \times U(1)^{k+1}$.  The $SU(n)^k$ theories are, in the IR, $k$ decoupled copies of SQCD, each with $N_f$ flavors.  The low-energy $SU(m)^2$ sector includes SQCD, with $N_f$ massless flavors, along with bifundamentals $F$ and $\tilde F$ coming from the adjoint $X$ of the original theory at the origin, with a low-energy superpotential $W_{low}\sim \text{Tr}(F\tilde F)^{(k+1)/2}$.    All other components from $X$ and $Y$ are either eaten in the Higgsing, or get a mass from the superpotential \dflatsup\ along the flat direction \Brodievev.  This low-energy theory is depicted in Fig. \ref{fig:FLATodd}, where as usual adjoints are arrows that start and end on the same node of the quiver diagram, and dotted adjoints depict those that get a mass term from the superpotential.  Brodie duality along this flat direction is then compatible with a duality in \rcite{ILS} (see Section 8 there) for the $SU(m)^2$ factor, and with Seiberg duality for the $SU(n)$ factors:  
	\eqna{
	\begin{array}{ccc}
	SU(2m+kn) & \overset{\langle X \rangle, \langle Y \rangle}{\longrightarrow}\ & SU(m)\times SU(m)\times SU(n)^k \times U(1)^{k+1} \\ \big\downarrow & & \big\downarrow \\ SU(3kN_f-(2m+kn)) & \longrightarrow & SU(kN_f-m)^2\times SU(N_f-n)^k\times U(1)^{k+1}
	\end{array}}[koddHiggsing]
where horizontal arrows are the flat direction \Brodievev\ and vertical arrows are the duality.

The low-energy $SU(m)^2$ theory has a further flat direction, where $F$ has non-zero expectation value, breaking to $SU(m)_D$ \rcite{ILS}.  The low-energy $SU(m)_D$ has an adjoint $\tilde{A}$, with superpotential $W \backsim \text{Tr} \tilde{A}^{\frac{k+1}{2}}$ corresponding to an $A_{(k-1)/2}$ theory with $2N_f$ flavors.  The duality of the low-energy $W_{A_{(k-1)/2}}$ theory along this flat direction then reduces to that of  \rcite{DKi}. We summarize these flat directions in Figure \ref{fig:FLATodd}. In sum, for $D_{\text{odd}}$, Brodie duality along the flat direction \Brodievev\ is nicely consistent with other dualities.  
\noindent

We now consider the analogous flat directions \Brodievev\ for the  puzzling $D_{\text{even}}$ cases, which again exist for $N_c=2m+kn$ and are parameterized by arbitrary $\lambda_x$. As in the $D_{\text{odd}}$ case, the gauge group is Higgsed $SU(2m+kn)\to SU(m)^2\times SU(n)^k\times U(1)^{k+1}$, where the $SU(m)^2$ and $SU(n)^k$ decouple from each other at low energies.  But the $D_\text{even}$ case differs from $D_\text{odd}$ in two respects.  First, the $SU(m)^2$ sector has massless bifundamentals $F$ and $\tilde F$, with $W_{low}(F\tilde F)=0$.  Similarly, the $SU(n)^k$ sector reduces at low-energy to $k/2$ decoupled copies of $SU(n)^2$
which each have, in addition to $N_f$ flavors, massless  bifundamentals with $W_{low}(F\tilde F)=0$.  For example, for $k=2$, $N_c=2(m+n)$, and expanding \dflatsup along the flat direction \Brodievev\ gives 
	\eqna{
	W_{SU(m)^2} & \supset \frac{t_2}{3}  \text{Tr}  A_{x,1}^3 +   \text{Tr}  A_{x,1}A_{y,1}^2  + 2 \lambda_x^{1/2}  \text{Tr}  A_{x,1}A_{y,1} - \lambda_x \text{Tr} A_{x,1}+ t_2 \tr A_{x,1} F_x\tilde{F}_x \\
	&+ (1\to 2, A_{y}\to -A_{y})\\
	W_{SU(n)^2} &\supset \frac{t_2}{3}  \text{Tr}  A_{x,3}^3 +\text{Tr} A_{x,3}A_{y,3}^2 +  \left(\frac{\lambda_x}{t_2}\right)^{1/k} \left(t_2  \text{Tr}A_{x,3}^2 + \text{Tr}A_{y,3}^2 \right) - \lambda_x \text{Tr} A_{x,3} \\
	&+ \frac{t_2}{3}  \text{Tr}  A_{x,4}^3 +\text{Tr} A_{x,4}A_{y,4}^2 -  \left(\frac{\lambda_x}{t_2}\right)^{1/k} \left(t_2  \text{Tr}A_{x,4}^2 + \text{Tr}A_{y,4}^2 \right)  - \lambda_x \text{Tr} A_{x,4} \\
	&+ \text{Tr} (A_{x,3}+A_{x,4})F_y\tilde{F}_y.
	}[supeven]
Subscripts $x,y$ refer to which $SU(2m+2n)$ adjoint $X,Y$ the field comes from, the $A_{1,2}$ are $SU(m)$ adjoints, and the $A_{3,4}$ are $SU(n)$ adjoints.  
Both of these IR superpotentials reduce to $W_{low}(F\tilde F)=0$ upon integrating out the massive adjoints.  The $SU(m)\times SU(m)$ theories with bifundamentals and $W_{low}(F\tilde F)=0$ do not have a known dual.  Indeed, they have a flat direction where $F$ gets an expectation value and Higgses $SU(m)\times SU(m)\to SU(m)_D$, where the low-energy $SU(m)_D$ is an $\widehat A$ theory, with massless adjoint ${\cal X}$ (coming from $\tilde F$) and $2N_f$ fundamentals, with $W_{low}({\cal X})=0$.

\begin{figure}[h!]
\centering
	\begin{subfigure}[h]{0.75\textwidth}
		\includegraphics[width=\textwidth]{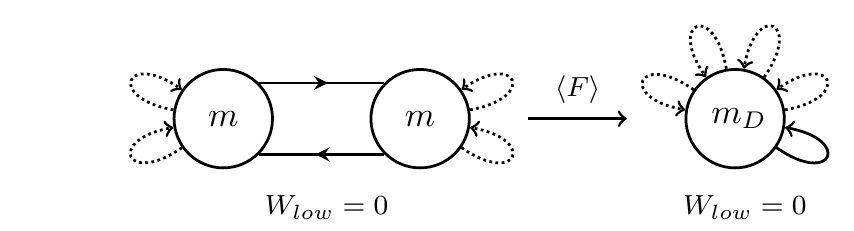}
		\caption{$SU(m)\times SU(m)$ sector, Higgses to $W_{\widehat{A}}$.}
		\label{}
	\end{subfigure}\\
	\begin{subfigure}[h]{0.75\textwidth}
		\includegraphics[width=\textwidth]{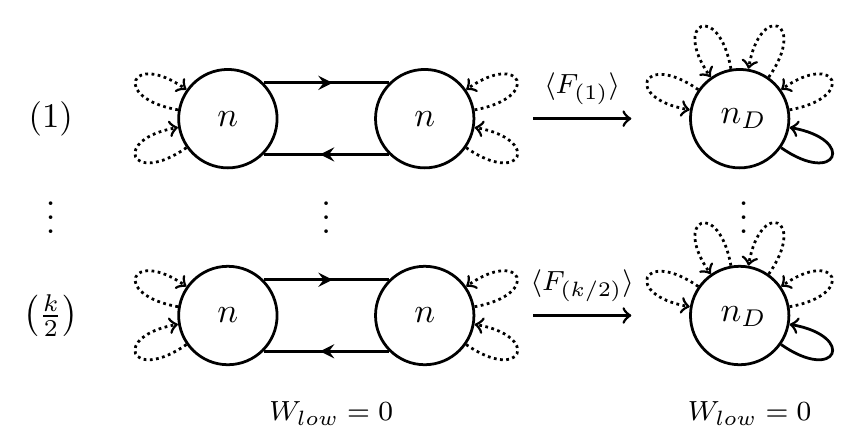}
		\caption{$SU(n)^k$ sector, Higgses to $k/2$ decoupled $W_{\widehat{A}}$ theories.}
		\label{}
	\end{subfigure}
\caption{Flat directions for $D_\text{even}$, with $N_c=2m+kn$. Again we integrate out massive fields (denoted by dotted lines), and those eaten by the Higgs mechanism (not shown).}
\label{fig:FLATeven}
\end{figure}

More generally, 
for even $k>2$, since $\omega = e^{2\pi i/k}$ in \Brodievev, there will be $k/2$ massless bifundamental pairs.  The low-energy $SU(n)^k$ theory then reduces to $k/2$ decoupled $SU(n)^2$ quiver gauge theories, where the $i$'th node couples to the $(k/2+i)$'th node via a pair of massless bifundamental fields. Each $SU(n)^2$ theory has a flat direction to an $SU(n)_D$ $\widehat A$ theory. The low-energy theories along these flat direction are as depicted in Figure \ref{fig:FLATeven}. 

The conclusion is that, for $D_\text{even}$, we end up with $(k/2+1)\ \widehat{A}$ theories corresponding to nodes with $2N_f$ flavors\footnote{For $N_c=2m$, there is a similar generalization of these flat directions parameterized by both $\lambda _x$ and $\lambda _y$, with $\langle X\rangle \propto \langle Y\rangle \propto \left(\begin{array}{cc} {\bf 1}_m & 0  \\ 0 & -{\bf 1}_m   \end{array}\right)$, which again leads to a low-energy $SU(m)_D$ $\widehat A$ theory with $2N_f$ flavors.}.  The $\widehat{A}$ theories along the flat direction are puzzling, as in Section \dline: we have not found a quantum mechanism for lifting these flat directions, and have not found how to make these flat directions compatible with Brodie's proposed duality.

\subsec{$D_{k+2}$ RG flows from relevant $\Delta W$ deformations}[Dflows]

In this subsection, we consider RG flows from the $W_{D_{k+2}}$ SCFTs upon deforming by relevant $\Delta W$.  As in the previous subsections, we find that cases involving only $D_\text{odd}$ are nicely compatible with the duality of \rcite{JB}, while those involving $D_\text{even}$ exhibit subtleties.  For simplicity, we mostly consider $U(N_c)$, with brief discussion of the more complicated $SU(N_c)$ version in Section \ref{sec:Dshift}. 

We begin with the class of $\Delta W$ deformation RG flows $D_{k+2}\to D_{k'+2}$, which is relevant for $k'<k$ (taking $x > x^{\text{min}}_{D_{k+2}}>x^{\text{min}}_{D_{k'+2}}$ as in \xmin):
	\eqn{
	W =  \text{Tr} XY^2 + \sum_{i=k'}^k \frac{t_i}{i+1} \text{Tr} X^{i+1},
	}[dtod]
which yields the F-terms
	\eqna{
	Y^2 + \sum_{i=k'}^k t_i X^i &= 0\\
	\{X,Y\} &=0.
	}[fterms]
The solution $X=Y=0$ corresponds to the $D_{k'+2}$ theory at the origin.  There are also $(k-k')$ 1d solutions with non-zero $X$-eigenvalue, corresponding to $A_1$'s. The representation theory of \fterms\ was discussed in \rcite{CachazoGH,twoadj}.  Taking $X$ and $Y$ to be matrices, it follows from \fterms\ that $X^2$ and $Y^2$ are Casimirs (proportional to the unit matrix),  so we may rewrite the first F-term as $(y^2 + Q_{\lfloor k/2\rfloor}(x^2)) {\bf 1} + P_{\lfloor (k-1)/2\rfloor}(x^2) X=0$, where the subscripts on $P$ and $Q$ denote the degrees of the polynomials in $x^2$.  There are 2d representations of the second F-term, taking $X =  x \sigma_2,\ Y = y \sigma_1$; then a non-zero solution for $X$ requires $ P_{\lfloor (k-1)/2\rfloor}(x^2)=0$. Hence, there are $\lfloor (k-1)/2\rfloor$ independent such solutions for $x^2$, and then $y^2$ is uniquely fixed\foot{$x\to -x$ or $y\to -y$ is a gauge rotation so does not give additional vacua.}.   If $X$ and $Y$ have $n_j$ copies of such a vacuum, where $j=1,\dots,  \lfloor (k-1)/2\rfloor$ labels the value of $x_j^2$, then the non-zero $X$ and $Y$ values break $SU(2n_j)\to SU(n_j)\times SU(n_j)\to SU(n_j)_D$, where the low-energy $SU(n_j)_D$ theory has $2N_f$ flavors.  Expanding $W(X,Y)$ in such vacua, the $X$ and $Y$ adjoints have mass terms and the low-energy theory is SQCD; we label such vacua as $A^{2d}_1$.  In sum, the $\Delta W$ deformation \dtod leads to vacua
	\eqna{
	D_{k+2}\  &\longrightarrow\  D_{k'+2} + (k-k') A_1 + \left(\left\lfloor \frac{k-1}{2}\right\rfloor - \left\lceil \frac{k'-1}{2}\right\rceil \right) A_1^{2d} }[dtodv]
The $N_c\times N_c$ matrices $X$ and $Y$ are decomposed into blocks, distributed among these vacua, with $n_0$ eigenvalues at the origin, $n_i$ at the $i$'th $A_1$ node, and $n_j^{2d}$  in the $j$'th $A_1^{2d}$ node, with $N_c=n_0+\sum _i n_i+2\sum _j n_j^{2d}$.  The gauge group is Higgsed in the electric and dual magnetic descriptions (for $x$ in the conformal window) as:
	\eqna{
	\begin{array}{clc}
	U(N_c)  & \overset{\langle X\rangle, \langle Y \rangle}{\longrightarrow}\ \ & {\displaystyle  U(n_0) \prod_{i=1}^{k-k'} U(n_i)  \prod_{j=1}^{\left\lfloor \frac{k-1}{2}\right\rfloor - \left\lceil \frac{k'-1}{2}\right\rceil}U(n^{2d}_j)} \\ 
 	\big\downarrow & &  \big\downarrow \\
	U(3kN_f-N_c)  & \ \longrightarrow &  {\displaystyle U(3k'N_f - n_0)\prod_{i=1}^{k-k'}SU(N_f-n_i) \prod_{j=1}^{\left\lfloor \frac{k-1}{2}\right\rfloor - \left\lceil \frac{k'-1}{2}\right\rceil } U(2N_f - n_j^{2d}) } \end{array}
	}[dkdkflow]
The down arrows are Brodie duality for the $D_{k+2}$ $U(N_c)$ theory in the UV, and Brodie or Seiberg duality for each approximately decoupled low-energy gauge group factor in the IR. Comparing the UV (LHS) and the IR (RHS) of the dual theories in the lower row of \dkdkflow, the IR theory only properly matches the dual Higgsing pattern of the UV theory if $\tilde N_c=\tilde n_0+\sum _i \tilde n_i+2\sum _j \tilde n_j^{2d}$.  This equality holds if and only if $k$ and $k'$ are both odd; this is a non-trivial check of Brodie duality for $D_\text{odd}\to D_\text{odd'}$.  But if either $k$ or $k'$ is even, there is a mismatch of $2 N_f$ between $\tilde{N}_c = 3kN_f-N_c$ on the LHS and its IR decomposition on the RHS of \dkdkflow, and
 a mismatch of $4N_f$ if both $k$ and $k'$ are even. 

We now consider the RG flow $D_{k+2}\to A_{k'}$, by adding $\Delta W= \frac{m_Y}{2} \text{Tr} Y^2$ to the superpotential in \dtod. There is then a low-energy $A_{k'}$ theory at the origin, $X=Y=0$, along with $(k-k')$ $A_1$'s corresponding to the 1d solutions of the vacuum equations with eigenvalues $y=0,\ \sum_{i=0}^{k-k'} t_{i+k'} x^i =0$, along with two more $A_1$ theories at $y=\pm\sqrt{- \sum_{i=k'}^k t_i x^i},\ x=-\frac{m_Y}{2}$.  As always, these 1d solutions of the $F$-term equations match the rank of the ADE group: $k+2$ in the UV matches the IR sum $k'+(k-k')+2$.  In addition, there are 2d representations of the $D$- and $F$-terms, with Casimirs $Y^2 = y^2 {\bf 1}$ and $\sum_{i=k'}^k t_i X^i = f(x) {\bf 1}$.  The 2d vacua may thus be parameterized as $X = -\frac{v}{2} {\bf 1} + x_1 \sigma_1,\ Y = y\sigma_3$, and the $F$-terms have $\lfloor (k-1)/2\rfloor$ solutions for $x_1$, each of which determines $f(x_1)$ and specifies the 2d vacuum.   In each such vacuum, the low-energy theory is SQCD (both $X$ and $Y$ have mass terms) with the $X$ and $Y$ expectation values breaking $SU(2n_j^{2d})\to SU(n_j)\times SU(n_j)\to SU(n_j)_D$, with $2N_f$ flavors in the low-energy theory.  In sum, the full (classical) structure of the vacua from such deformations is 
	\eqna{
	D_{k+2} &\longrightarrow A_{k'} +(k-k'+2) A_1 +  \left\lfloor \frac{k-1}{2} \right\rfloor A_1^{2d}.
	}[dtoa]
Taking $N_c=n_0+\sum _{i=1}^{k-k'+2}+\sum _{j=1}^{\lfloor (k-1)/2\rfloor }2n_j^{2d}$, the deformation results in the following Higgsing in the electric and magnetic descriptions: 
	\eqna{
	\begin{array}{clc}
	U(N_c)  & \overset{\langle X \rangle, \langle Y \rangle}{\longrightarrow}\   & {\displaystyle U(n_0)\times \prod_{i=1}^{k-k'+2} U(n_i) \times \prod_{j=1}^{\lfloor \frac{k-1}{2}\rfloor }U(n^{2d}_j) }\\ 
	\big\downarrow   & & \big\downarrow \\
	U(3kN_f-N_c) &\  \longrightarrow  & {\displaystyle U(k'N_f - n_0)\prod_{i=1}^{k-k'+2}U(N_f-n_i)\prod_{j=1}^{\lfloor \frac{k-1}{2}\rfloor } U(2N_f - n_j^{2d})} \end{array}.
	}[dkAkflow]
Again, the down arrows are duality in the UV theory on the LHS, and in each of the low-energy decoupled IR theories on the RHS.  Again, for odd $k$ the UV and the IR groups properly fit together, while for even $k$ there is a mismatch in the dual gauge group of  $2N_f$.

In summary, whenever the RG flows only involve $D_\text{odd}$, there is a successful, non-trivial check that the deformation maps properly between the UV and IR theories.  On the other hand, whenever we flow to/from a $D_{\text{even}}$ theory, there is a mismatch in the dual gauge groups pre and post deformation.  An especially peculiar mismatch arises if we flow through an intermediate $D_{k'=\text{even}}$ theory, first deforming by $\sum_{i=k'}^{k-1} t_i X^{i+1}$ as in \dtodv, and then deforming by $\frac{v}{2} \text{Tr} Y^2$ as in \dtoa, which gives
	\eqna{D_{k+2}\  \longrightarrow\  A_{k'} + (k-k'+2) A_1 + \left(\left\lfloor \frac{k-1}{2}\right\rfloor - \left\lceil \frac{k'-1}{2}\right\rceil +\left\lfloor \frac{k'-1}{2}\right\rfloor \right) A_1^{2d}.}[Dflowtwo]
For $k'$ even, $\lfloor \frac{k'-1}{2}\rfloor- \lceil \frac{k'-1}{2}\rceil =-1$, and the number of 2d vacua in \Dflowtwo\  differs from that in \dtoa\ for flowing directly with both  $\sum_{i=k'}^{k-1} t_i X^{i+1}$ and $\frac{v}{2} \text{Tr} Y^2$ deformations together.  Perhaps the conjectured quantum truncation of the chiral ring for $D_\text{even}$ eliminates these puzzling mismatches in the higher dimensional representations for these flows, but we have not yet succeeded in showing how. We leave this as a challenge for future understanding.

\subsec{The $SU(N_c)$ version of the RG flows} \label{sec:Dshift}

The above analysis was for $U(N_c)$.  To adapt it for $SU(N_c)$, we write $X_{U(N_c)}=X_{SU(N_c)}+X_0{\bf 1}_{N_c}$, where $\text{Tr}X_{SU(N_c)}=0$, and likewise for $Y$, and can eliminate the unwanted $X_0$ and $Y_0$ fields via Lagrange multipliers, as in Section \ref{sec:su}.  The complication is that if we want to keep the enhanced $D_{k'+2}$ or $A_{k'}$ singularities as in \dtodv\ or \dtoa, we need to add lower order $\Delta W$ terms, beyond those already present for the $U(N_c)$ version of the RG flows.  These extra terms are needed in order to re-tune, to zero, the corresponding $\Delta W$ relevant deformations which would be generated by adding the Lagrange multiplier constraint terms, and which would generically further deform the RG flow to merely multiple $A_1$ vacua.  For flows starting at $D_{k+2}$ as in \WDk, the needed deformations  are included in
	\eqna{
	\Delta W \subset \sum_{i=1}^{k-1} \frac{t_i}{i+1} \text{Tr} X^{i+1} + \sum_{i=0}^{\left\lfloor \frac{k}{2} \right\rfloor} \frac{u_i}{i+1}\text{Tr} X^{i+1}Y + \frac{m_Y}{2} \text{Tr}Y^2 - \lambda_x \text{Tr} X - \lambda_y \text{Tr} Y.
	}[defs]

For generic couplings in \defs, the RG flow leads to vacua as
\eqna{D_{k+2}\to (k+2)A_1+\left\lfloor \frac{k-1}{2}\right\rfloor A_1^{2d},}
which is the same for $SU(N_c)$ and $U(N_c)$. One can now tune the couplings in \defs\ to enhance to an $A_{k'}$ or $D_{k'+2}$ singularity, and then the flow involves Higgsing as in e.g. \dkdkflow, but with all $U(N)$ factors replaced with $SU(N)$.  The tuning shifts of the couplings in \defs\ are complicated, and depend on how many eigenvalues $n_0$ are in the enhanced $D_{k'+2}$ or $A_{k'}$ vacua.  We have verified that, despite these technical complications, the vacuum structure is qualitatively similar to that of the $U(N_c)$ case, replacing $U(n)\to SU(n)$ everywhere in Section \Dflows.  

Interestingly, there can be several options in performing the wanted shift, and these can result in different Casimirs along the flow. We illustrate this for the example $D_5\to D_3$, and note that there are similar versions for other $D$ flows.  The first way to enhance to $D_3$ is via a tuned addition of the $\{ m_Y,\lambda_x\}$ deformations to \dtod, where the needed shift of these couplings depends on the $\{ t_3,t_2,t_1\}$ couplings in \dtod, as well as the multiplicities of the eigenvalues in the vacua.  The Casimirs along the flow are then $Y^2$ and $t_3X^3+t_2X^2+t_1X$.  Much as in \dtodv, we indeed find one $A_1^{2d}$ vacuum.  Another option for $D_5\to D_3+\dots$ is to add only the  $\frac{u_1}{2} \text{Tr} X^2 Y$ deformation in \defs, with the other $\Delta W$ couplings set to zero. Then $X^2$ and $Y^2$ are Casimirs, but $X$ and $Y$ no longer anticommute as they did in the $U(N_c)$ case, and so a 2d solution is now of the form $X = x_1 \sigma_1 - ix_3\sigma_3,Y = y_1 \sigma_1 + i y_3\sigma_3$. We again find one 2d representation of the $F$- and $D$-terms, which reduces to the $U(N_c)$ 2d solution as $u_1\to 0$.  Different sets of lower order deformations in the chiral ring lead to different Casimirs along the flow, but nevertheless non-trivially give the same counting for the higher dimensional vacua.

\subsec{The $D_\text{odd}\to D_\text{even}$ RG flow and the hypothetical $D_\text{even}'$ theory}[dcomm]%

As discussed in the previous subsections, the $D_\text{even}$ theories have some puzzles, whereas the $D_\text{odd}$ theories appear to be under control.  This suggests trying to understand the $D_\text{even}$ theories via RG flows from the understood UV case: $D_\text{odd}\to D_\text{even}$.  Indeed, the idea of embedding $D_\text{even}$ in $D_\text{odd}$ was the basis for the original conjecture  \rcite{JB}\ that quantum effects somehow make the  troubling $D_\text{even}$ theories similar to the nice $D_\text{odd}$ theories.  In this subsection, we examine the  $D_\text{odd}\to D_\text{even}$ RG flow more carefully, and note that this flow has its own subtleties.

As seen in \dtodv, the $\Delta W$ RG flow from $D_{k+2}\to D_{k'+2}$ comes with jumping number of $A_1^{2d}$ representations, from the floor and ceiling functions, which is only straightforward for the $D_\text{odd}\to D_\text{odd'}$ cases.  We here further discuss the relation and difference between $D_\text{odd}\to D_\text{odd}$ vs  $D_\text{odd}\to D_\text{even}$.   Consider starting from the $D_{k+2}$ SCFT, with $k$ odd, and deforming by $\Delta W$.  To simplify the discussion, we consider $U(N_c)$ (as opposed to $SU(N_c)$) and start with the $\Delta W$ deformation considered in \dtodv with $k'=k-2$: $D_{k+2}\to D_k + 2A_1 + A_1^{2d}$. The low-energy $D_k$ theory is at $X=Y=0$, the $2A_1$ theories are at $X$ having eigenvalues $x_\pm$ with $Y=0$, and the $A_1^{2d}$ theory has $(X,Y)$ values at $(x_{2d}, y_{2d})$ given by:
	\eqna{
	(x,y) = \left\{ 
	\begin{array}{cc}
	(0,0) &  \\
	(x_{\pm},0) & t_kx_{\pm}^2 + t_{k-1} x_{\pm} + t_{k-2} = 0,  \\
	(x_{2d},y_{2d}) & t_k (x_{2d})^2 + t_{k-2} = 0,\quad (y_{2d})^2 + t_{k-1} (x_{2d})^{k-1} = 0.
	\end{array} \right.
	}[ds]
If we start at the $D_{k+2}$ theory (as opposed to $\widehat D$), we can set $t_k=1$, and $t_{k-1}$ and $t_{k-2}$ are the $\Delta W$ deformation parameters.

\begin{figure}[h!]
\centering
	\begin{subfigure}[h]{0.59\textwidth}
		\includegraphics[width=\textwidth]{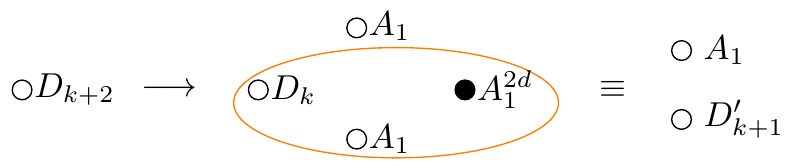}
		\caption{The 2d vacuum and an $A_1$ both collapse to the origin.}
		\label{fig:d1}
	\end{subfigure}\\
	\begin{subfigure}[h]{0.59\textwidth}
		\includegraphics[width=\textwidth]{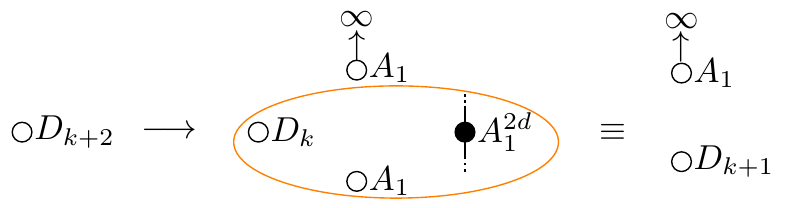}
		\caption{One $A_1$ goes to the origin while the other goes off to infinity, and the 2d vacuum becomes a degenerate line of 2d representations as in \FDline.}
		\label{fig:d2}
	\end{subfigure}
\caption{We enhance the $k$ odd $D_k$ singularity to a $D_{k+1}$ singularity in one of two ways. \label{fig:denhancement}}
\end{figure}

We now try to tune the superpotential couplings to collide the $D_k$ singularity with an $A_1$ singularity, to get an enhanced $D_{k+1}$ singularity.  This can be accomplished by tuning $t_{k-2}\to 0$ in \ds, which brings one of the $A_1$ singularities ($x_+$ or $x_-$) to the origin.  Note that $t_{k-2}\to 0$ also brings $x_{2d}$ and $y_{2d}$ to the origin.  We denote this enhancement as $D_k+A_1+A_1^{2d}\to D_{k+1}'$, where the prime distinguishes the theory from the even $D_{k+1}$ theory that one would obtain by flowing directly from the $\widehat D$ theory.  We can formally obtain that latter theory, $D_{k+1}$, directly from the $\widehat D$ fixed point, by taking $t_k\to 0$ along with $t_{k-2}\to 0$ in \ds; this brings one of the $x_\pm$ to the origin and the other to infinity, and then the last equation in \ds\ gives the line of $A_1^{2d}$ solutions \FDline where $D_\text{even}\to \widehat A$, since \ds\ is satisfied for all $x_{2d}$ when $t_k=0$.   The two procedures are indicated in the Figure \ref{fig:denhancement}.

The two procedures suggest that perhaps there are actually two types of $D_\text{even}$ theories.  One is the $D_\text{even}'$ theory of Figure \ref{fig:d1}, which can actually be obtained from the RG flow $D_\text{odd}\to D'_\text{even}$, and which therefore inherits the simpler properties of $D_\text{odd}$.  The other is the mysterious $D_\text{even}$ theory of Figure \ref{fig:d2}, which actually is not obtained from RG flow from $D_\text{odd}$, but instead only from $\widehat D\to D_\text{even}$, since it requires $t_k=0$ and the $D_\text{odd}$ theory had $t_k=1$.  The latter, $D_\text{even}$ theory has the puzzles, discussed in the previous subsections, associated with the $D_\text{even}\to \widehat A$ moduli space of vacua and the non-truncated chiral ring.   

We have thus considered the possibility that Brodie duality actually only applies to the simpler $D_\text{even}'$ theory, which inherits the truncated chiral ring from $D_\text{odd}$, and does not apply to the $D_\text{even}$ theory.   However, this scenario also has challenges.  If we take seriously the idea that a $D_{k+1}'$ (for $k$ odd) theory is made by bringing together $D_{k}+A_1 + A_1^{2d}$, this seems to suggest that the chiral ring of the $D_{k+1}'$ theory contains  $(3k-1)N_f^2$ mesonic operators, where the $-N_f^2$ are those in the $A_1$ singularity, which decouples from $D_{k+1}'$ in the IR.  On the other hand, assuming that Brodie duality applies to $D_{k+1}'$, we would have expected $3(k-1)N_f^2$ mesonic operators.  The $D_{k+1}'$ theory has an extra $2N_f^2$ mesonic operators.  Perhaps then, in collapsing the $A_1$ and $A_1^{2d}$ theories to the $D_{k}$ theory at the origin, a slightly modified version of Brodie duality applies, with  $\alpha_{D_{k+2}'} = 3k+2$. We have also tried to cure the apparent $a$-theorem violations of Section \athm by adding the $2N_f^2$ mesons to the UV $D_\text{even}$ theory.  But the results did not look promising: the extra operators seem to become free at too large $x$ to cure the apparent wrong sign of $\Delta a$.  It is still possible that some modified version of Brodie duality resolves these puzzles, and we invite the interested reader to try.

\newsec{The $W_{E_7}$ fixed point and flows} \label{sec:E}

The $W_{E_7}$ SCFT arises as the IR limit of a relevant superpotential deformation to the $\widehat{E}$ SCFT, with corresponding superpotential 
	\eqn{
	W_{E_7} = \frac{1}{3} \tr Y^3 + s_1\tr Y X^3.
	}[Esup]
The $\tr Y X^3$ term is a relevant deformation to the $\widehat{E}$ fixed point for $x>x_{E_7}^{\text{min}} \approx 4.12$, where $x_{E_7}^{\text{min}}$ was determined via $a$-maximization in \rcite{twoadj}; here we will assume that $x>x_{E_7}^{\text{min}}$. 

The F-terms of the undeformed $E_{7}$ superpotential in \Esup\ are given by
	\eqna{
	Y^2 + s_1 X^3 &= 0,\\ 
	X^2Y+XYX+YX^2 &= 0,
	}[EFterms]
from which it follows that the chiral ring does not truncate classically. We may write the generators of the classical chiral ring in a basis 
	\eqna{
	\Theta_{(1,n)} &=X^n,\\
	\Theta_{(2,n)} &= Y X^n,\\
	\Theta_{(3,n)} &= XYX^n,\\
	\Theta_{(4,n)} &= YXYX^n;\ \ \ n=0,1,...
	}[Egen]

\subsec{Proposed dualities for $W_{E_7}$ \rcite{DKJLyqa}}[pde]

It was noted in \rcite{twoadj} that the $W_{E_6}$ apparently violates the $a$-theorem condition \aderivneg\ for $x\gtrsim 13.8$, and thus some new dynamics must enter for $x\sim 13.8$ (or less) to ensure that the $a$-theorem is satisfied.  In \rcite{DKJLyqa}, it was pointed out that for the $W_{E_7}$ theories the condition \aderivneg\ is violated for $x\gtrsim 27$, so some new dynamics is needed there, or at smaller $x$.  The dual theory proposed in \rcite{DKJLyqa} resolves this apparent $a$-theorem violation, since it implies different IR phases for larger $x\gtrsim 26.11$ \rcite{DKJLyqa}. The duality of  \rcite{DKJLyqa} requires that the chiral ring  truncates, similar to the conjecture in \rcite{JB} for $D_{\text{even}}$, as 
	\eqn{
	YX^6+b XYX^5=0 \quad \hbox{in the chiral ring}
	}[pc]
for some constant $b$. It is not yet known if the proposed quantum constraint \pc\ is correct, or how it arises.  Imposing \pc, the chiral ring of the electric theory is truncated to 30 independent generators  (listed for reference in Appendix \ref{sec:cr}). The resulting IR dual description of the $E_7$ fixed point has gauge group $SU(\alpha_{E_7}N_f-N_c)$ with $\alpha_{E_7}=30$, and the usual duality map reviewed in Section \matrix. The dual theory has superpotential\foot{As in \rcite{DKJLyqa}, we scale the factors of $\mu$ to unity.} \rcite{DKJLyqa}
	\eqn{
	W_{E_7}^{mag} \backsim  \frac{1}{3} \tr \hat{Y}^3 + \hat{s}_1\tr \hat{Y} \hat{X}^3 + \sum_{j=1}^{30} M_{j} \tilde{q} \Theta_{30-j} (\hat{X},\hat{Y}) q.
	}[WEmag]  
In addition to the usual tests of duality---'t Hooft anomaly matching, that the charge assignment for the magnetic fields under the global symmetry is consistent with the duality map---it was verified in  \rcite{Kutasov:2014wwa} that the superconformal index of the dual theories agrees, at least in the Veneziano limit (away from that limit, the duality and agreement of their superconformal indices suggests new mathematical identities). 

As we discuss in the following subsections, we find similar puzzles for the $E_7$ theories as with the $D_{\text{even}}$ theories.  In the following, we mirror our analysis of the $W_{D_{k+2}}$ theories for $W_{E_7}$; as such, we will be brief when analysis or discussion is similar to what has already been discussed in Section \ref{sec:D}. Much as we found for $D_{\text{even}}$, we fail to find evidence for this truncation, and point out additional hurdles for the conjectured duality.

\subsec{Matrix-related flat directions at the origin}[ematrix]

We consider the moduli space of vacuum solutions of \EFterms with $D$-term constraints \Dterms, setting $Q=\tilde{Q}=0$. The only 1d solution corresponds to the $E_7$ singularity at the origin. The first $F$-term in \EFterms shows that $Y^2$ and $X^3$ are Casimirs, yielding Casimir conditions $X^3=x^3{\bf 1}_d$, and $Y^2=y^2{\bf 1}_d$ for a $d$-dimensional representation. There is a line of $d=2$ solutions to these conditions analogous to \FDline,
	\eqna{
	X= x \left(\begin{array}{cc} \omega  & 0\\ 0&  \omega^2  \end{array} \right),\quad Y = y \sigma_1\\
	y^2 + s_1 x^3 = 0.
	}[Eline]
for $ \omega = e^{2\pi i/3}$. As $X$ and $Y$ are not traceless, this flat direction is present for only $U(N_c)$\footnote{For special cases of \Eline there will be $SU(N_c)$ flat directions; for example, when there are equal multiplicities of $X$, $\omega X$, and $\omega^2 X $ along the line given in \Eline. In that case, one could check the proposed $SU(N_c)$ duality along the corresponding flat directions.}.

In general, $E_7$ has vacua with multiple copies of the solution \Eline, with the remaining eigenvalues of $X$ and $Y$ at the origin, giving a moduli space of supersymmetric vacua labeled by $y_i^2$ and $x_i^3$ satisfying \Eline, for $i=1,\dots, \lfloor N_c/2\rfloor$. These vacua Higgs the gauge group in a way that turns out to be analogous to the  $D_{\text{even}}$ case discussed in \dline. In particular, for $N_c=2n$ with $n$ copies of the 2d vacuum \FDline and unequal expectation values of the $y_i^2,x_i^3$, the resulting breaking pattern is $U(2n) \to U(n)_D \to U(1)^{n}$.  In summary, much as in \DevenAhat, there is a (classical) flat direction:
\eqn{E_7 \to \widehat A, \qquad\hbox{with} \quad U(N_c)\to U(\lfloor N_c/2\rfloor)_D \quad\hbox{and}\quad  N_f^{low}=2N_f,}[EsevenAhat]
so $x^{low}=N_c^{low}/N_f^{low}=(N_c/2)/(2N_f)=x/4$. If we assume that Kutasov-Lin's duality \rcite{DKJLyqa} holds, then as in the $D_{\text{even}}$ case we are led to a puzzle similar to that of the $D_\text{even}$ theories: the moduli spaces of the electric theory and its dual differ, and the low-energy theories on the flat directions of the two conjectured duals, $SU(N_c/2)_{\widehat A}$ and $SU(15N_f-N_c/2)$, are not clearly related. This flat direction is related to the classical nontruncation of the $E_7$ chiral ring, and again provides us with a way to sharpen the puzzle of how the truncation occurs by asking what lifts the flat direction.

Independent of the conjectured duality \rcite{DKJLyqa}, the deformation \EsevenAhat\  seemingly violates the $a$-theorem \atheorem for sufficiently large $x$.  As in Section \athm, we compute $a_{UV}(x)$ for the $W_{E_{7}}$ theory, with gauge group $U(N_c)$ and $N_f$ flavors, as in \rcite{twoadj}.  Likewise, $a_{IR}(x)$ for the $\widehat{A}$ theory, with gauge group $U(N_c/2)$ and $2N_f$ flavors, is computed as in \rcite{KPS}. We include the effects of all mesons hitting the unitarity bound assuming that the chiral ring is quantumly truncated, such that all the operators listed in Appendix \ref{sec:cr} are taken into account, and work in the Veneziano limit. We plot until the bottom of the conformal window---which occurs before the electric $E_7$ theory's stability bound, $x<30$ as predicted by duality---such that we expect the $a$-theorem to hold in the whole range plotted. 

	\begin{figure}[h]
	\centering
		\includegraphics[width=0.5\textwidth]{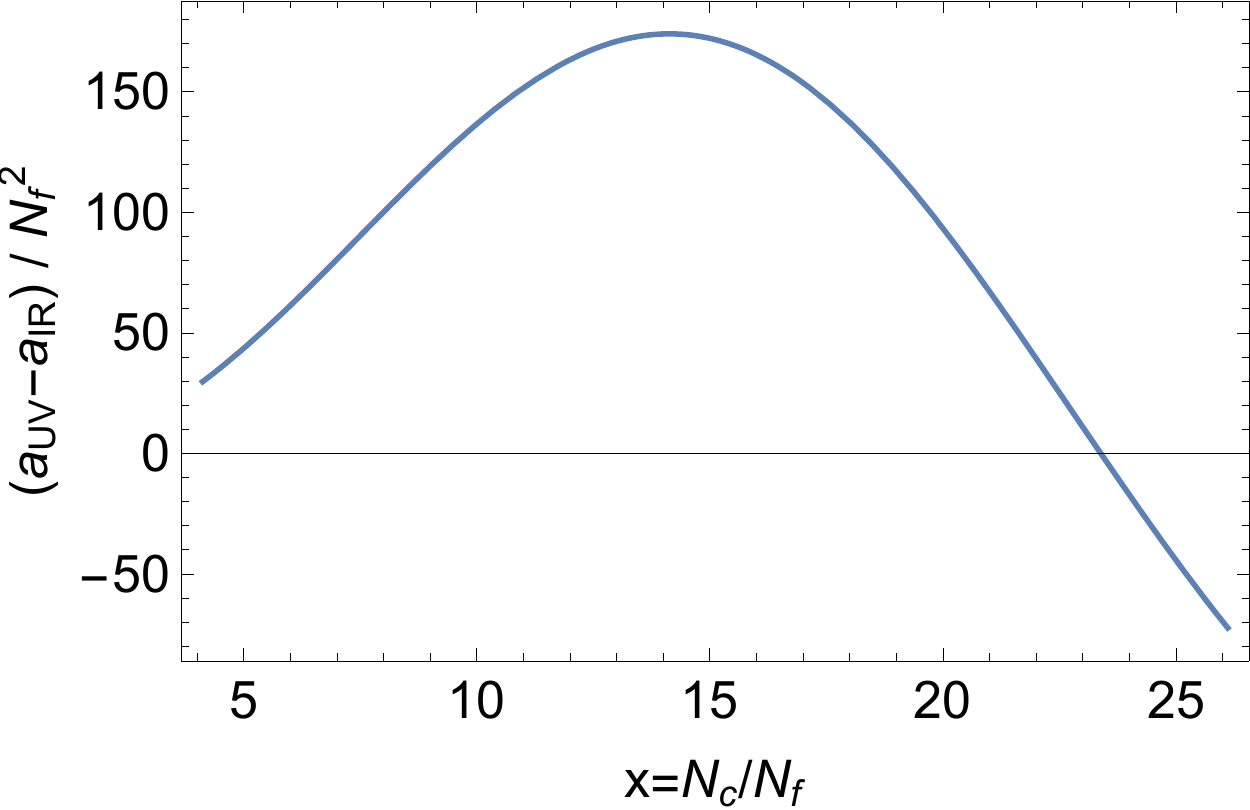}
		\caption{$(a_{UV}-a_{IR})/N_f^2$ for $W_{E_{7}}$ in the UV and $\widehat{A}$ in the IR. The  $E_7$ deformation term in the UV theory is relevant for $x\gtrsim 4.12$, while the corresponding term in Kutasov-Lin dual is relevant if $x\lesssim 26.11$. The $\hat{A}$ theory is UV-free in this whole range.}
		\label{fig:Ea}
	\end{figure}

As seen Figure \ref{fig:Ea}, this flat direction seems to violate the $a$-theorem in the conformal window for $x\gtrsim 23.39$. Unlike the $D_{\text{even}}$ case, this violation occurs for $x$ larger than the value where the mesons removed by the proposed quantum constraint \pc\ would hit the unitarity bound and become free; the first such meson that would be nonzero involves the operator $YX^6$, which would become free at $x=21$. To understand the effect that these would-be mesons would have on the computation of $a$ for this flat direction, we have performed the same check as in Figure \ref{fig:Ea}, but without imposing the proposed constraint. It turns out that this is not enough; the effect of including these operators in the ring is only to push the range of the apparent $a$-theorem violation to $x\gtrsim 23.44$.

The apparent violation of the $a$-theorem for these flat directions must of course be somehow resolved.  As in the discussion in Section \athm, either these flat directions are lifted in a way we don't understand, or some additional degrees of freedom make the calculation of $a$ incorrect---perhaps in the UV $W_{E_7}$ theory. The arguments made in Section \qflat would also apply here, and suggest that the former is not the solution. Since the calculation of $a$ in Figure \ref{fig:Ea} already took into account the proposed $W_{E_7}$ duality, we are left with a puzzle.

\subsec{$SU(N_c)$-specific (as opposed to $U(N_c))$ flat directions}[esun]

We now study $SU(N_c)$ flat directions of the $W_{E_7}$ theory, imposing the tracelessness of the adjoints with Lagrange multipliers $\lambda_x,\lambda_y$:
	\eqn{
	W_{E_7} =  \frac{1}{3} \tr Y^3 + s_1\tr Y X^3 - \lambda_x \tr X - \lambda_y \tr Y.
	}[Eflatsup]
When $N_c=2m+3n$ for $m,n$ integers, there is a flat direction labeled by $\lambda_y$
	\eqna{
	\langle X \rangle = \left(\frac{\lambda_y}{s_1}\right)^{\frac{1}{3}} &\left( \begin{array}{ccccc}   {\bf 0}_m  & & & & \\  & {\bf 0}_m   & & & \\ & & \omega {\bf 1}_n  & &  \\ & & & \omega^2 {\bf 1}_n & \\ & & & & \omega^3 {\bf 1}_n \end{array}\right),\\
	\langle Y \rangle = \left(\lambda_y\right)^{\frac{1}{2}} &\left( \begin{array}{ccccc}  - {\bf 1}_m & & & & \\  &   {\bf 1}_m   & & & \\ & &   {\bf 0}_n   & &  \\ & & & {\bf 0}_n & \\ & & & & {\bf 0}_n\end{array}\right),
	}[Evev]
where $\omega=e^{2\pi i/3}$ and off-diagonals are zero. \Evev is the special case of $k=3$ in \Brodievev. 

\begin{figure}[t]
	\centering
		\begin{subfigure}[h]{0.69\textwidth}
			\includegraphics[width=\textwidth]{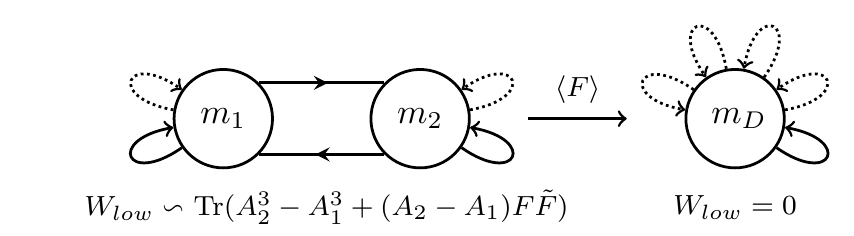}
			\caption{$SU(m)_1\times SU(m)_2$ sector, Higgses to $W_{\widehat A}$.}
			\label{fig:2m}
		\end{subfigure}\\
		\begin{subfigure}[h]{1\textwidth}
			\includegraphics[width=\textwidth]{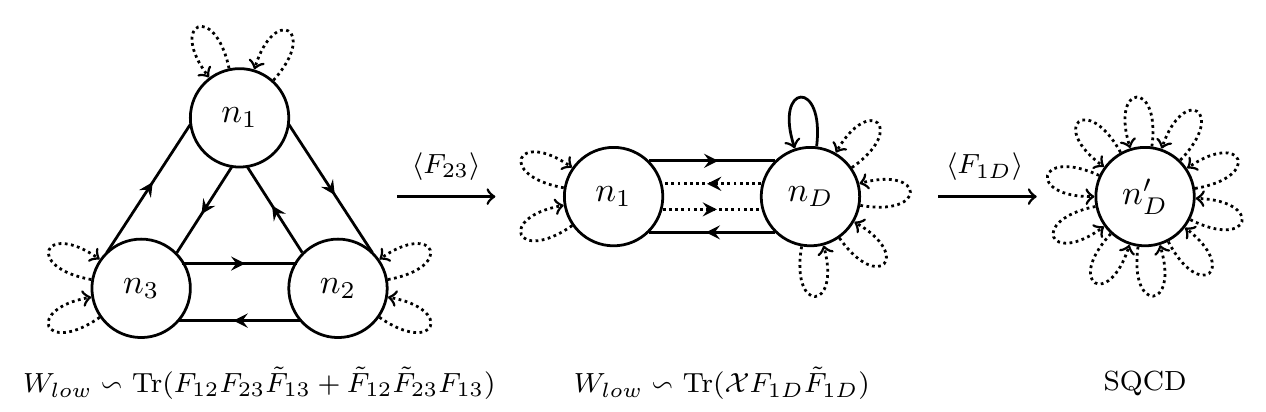}
			\caption{$SU(n)_1\times SU(n)_2\times SU(n)_3$ sector, Higgses to SQCD.}
			\label{fig:3n}
		\end{subfigure}
	\caption{Flat directions for $E_7$, $N_c=2m+3n$, integrating out massive fields (denoted by dotted lines) and fields eaten by the Higgs mechanism (not shown). The subscripts label the gauge groups and their matter. \label{fig:FLATe}}
	\end{figure}

Along this flat direction, the gauge group is Higgsed $SU(2m+3n) \to SU(m)^2\times SU(n)^3\times U(1)^4$. The low-energy $SU(m)^2$ sector includes $N_f$ massless flavors, along with bifundamentals $F,\tilde F$ and adjoints $A_1,A_2$ coming from the adjoint $X$ of the original theory at the origin, with a low-energy superpotential that is cubic in the massless fields (written in Figure \ref{fig:2m}). Thus, each $SU(m)$ node corresponds to a $W_{A_2}$ theory plus extra flavors from the bifundamentals. The low-energy $SU(n)^3$ sector includes $N_f$ massless flavors along with three pairs of bifundamentals $F_{12},F_{23},F_{13}$, and their conjugates, coming from the adjoint $Y$ of the original theory at the origin. There is an IR superpotential for these fields $W_{low} \backsim \text{Tr} (F_{12} F_{23} \tilde{F}_{13} + \tilde{F}_{12} \tilde{F}_{23}F_{13})$, which corresponds to making a loop around the quiver diagram shown in Figure \ref{fig:3n}. All other components from $X$ and $Y$ are either eaten in the Higgsing, or get a mass from the superpotential \Eflatsup, such that the $SU(m)^2$ and $SU(n)^3$ sectors decouple from each other at low energies. These low-energy theories are summarized in the left-most quiver diagrams in Figure \ref{fig:FLATe}.

We can then go along a further flat direction of the low-energy $SU(m)^2$ theory, where we give an arbitrary vev to the massless $F$, such that $SU(m)_1\times SU(m)_2$ breaks to the diagonal subgroup $SU(m)_D$. The low-energy $SU(m)_D$ has an adjoint that remains massless, and IR superpotential $W_{low}=0$ from integrating out the massive fields, such that this node corresponds to an $\widehat{A}$ theory with $2N_f$ massless flavors. This IR theory is depicted on the RHS of Figure \ref{fig:2m}.

The low-energy $SU(n)^3$ sector has a similar series of flat directions, where one of the massless bifundamentals has non-zero expectation value, depicted by the arrows in Figure \ref{fig:3n}. For example, giving a vev first to $F_{23}$ breaks $SU(n)^3\to SU(n)_D\times SU(n)$, resulting in an IR theory with one massless adjoint $\cal{X}$ charged under $SU(n)_D$ and one massless bifundamental pair coming from $F_{13},\tilde{F}_{12}$. Identifying the indices appropriately, these massless fields have an IR superpotential $W_{low}\backsim \text{Tr} ({\cal X} F_{1D} \tilde{F}_{1D})$. At this stage there is another flat direction where $F_{1D}$ has a non-zero expectation value, Higgsing $SU(n)\times SU(n)_D\to SU(n)_D$, where the remaining node corresponds to SQCD with $3N_f$ massless flavors. 

Again, it is not known if the $\widehat A$ theory in \ref{fig:2m} at low-energy has a dual.  On the other hand, the low-energy SQCD theory in \ref{fig:3n} has a dual given by Seiberg duality. So we can consider, for example, $N_c=3n$ ($m=0$), which along the flat direction of \ref{fig:3n} breaks to $SU(n)$ SQCD with $3N_f$ flavors, and thus has a Seiberg dual with gauge group $SU(n)_D\to SU(3N_f-n)_D$. The dual theory must also have moduli corresponding to the flat direction of the electric theory.  Extrapolating back to the origin along these dual flat directions, to try to un-Higgs the $SU(3N_f-n)_D$ dual by reversing the process analogous to \ref{fig:3n}, suggests a dual gauge group of $SU(9N_f - 3n)$ on the magnetic side at the origin. This disagrees with the dual gauge group of Kutasov-Lin's, which maps $SU(3n)\to SU(30N_f - 3n)$.  That latter theory has a flat direction, corresponding to \ref{fig:3n} where instead $SU(30N_f-3n)\to SU(10N_f-n)_D$.

To summarize, these flat directions pose puzzles for the proposed $W_{E_7}$ duality, both in the $\widehat A$ theory of \ref{fig:2m}, and in the SQCD theory of \ref{fig:3n}. We have not found a quantum mechanism for lifting these flat directions.

\subsec{Case studies in $E_7$ RG flows from $\Delta W$ deformations} \label{sec:Eflows}

In this subsection, we deform the $W_{E_7}$ SCFT \Esup by several examples of relevant $\Delta W$ for cases where the resulting IR theory is under better control in terms of duality. We study how the vacuum structure matches between the UV and IR electric and magnetic descriptions, focusing on the apparent puzzles of the UV $E_7$ theory. 

These examples demonstrate several new features, as compared with the $A_k$ and $D_{k+2}$ series.  One difference is that the deformed chiral ring admits $d>2$ dimensional representations. Further, we explore cases in which enhancements of the singularities in the IR of an RG flow (via tuning couplings of the deformations) do not preserve the number of higher-dimensional vacua. Interestingly, for some RG flows the $SU(N_c)$ version of a flow with the same 1d vacuum structure as the corresponding $U(N_c)$ flow has a different set of higher-dimensional vacua. Furthermore, we explicitly construct some RG flows for which the $\Delta W$ deformations are not apparently relevant.

\subsubsec{$E_7\to A_2$: 3d vacua}

We begin with the RG flow $E_7\to A_2$ flow for gauge group $U(N_c)$, taking $x>x^{\text{min}}_{E_7}$:
	\eqn{
	W = \frac{1}{3} \text{Tr} Y^3 +  s_1 \text{Tr} Y X^3 + \frac{t_1}{2} \text{Tr}X^2,
	}[etoa]
which yields the $F$-terms
	\eqna{
	Y^2 + s_1 X^3 &=0\\
	s_1(Y X^2 + X Y X + X^2 Y) + t_1 X &=0.
	}[etoafterms]
There are seven 1d solutions to \etoafterms: two coincident at $X=Y=0$, corresponding to the $A_2$ theory, and five solutions with nonzero $X$ and $Y$ eigenvalues, corresponding to $A_1$ theories; as always, the 1d solutions correspond, as in Arnold's ADE singularity resolutions, to adjoint Higgsing of the $G=ADE$, and preserving $r_G$.  Taking $X$ and $Y$ to be matrices, it follows from \etoafterms that $X^3\backsim Y^2$ are Casimirs along the flow, so that we may write $X^3=x^3 {\bf 1}_d$ and $Y^2=y^2 {\bf 1}_d$ for a $d$-dimensional representation. There is a 2d as well as a 3d representation that solve the $F$-terms, $D$-terms \Dterms, and Casimir conditions, 
	\eqna{
	X_{2d} &= \frac{1}{2} \sqrt{\frac{|t_1|^2}{|s_1|}}(\sigma_1 + i \sigma_3),\quad Y_{2d} = \frac{1}{2} \sqrt{\frac{|s_1|t_1^{3/2}}{s_1^2 (t_1^*)^{1/2}}} (\sigma_1 - i\sigma_3)\\
	X_{3d} &= \sqrt{\frac{t_1}{2s_1}} \left( \begin{array}{ccc} 0 & 1 & 1 \\ 0 & 0 & 0 \\ 0 & 0 & 0 \end{array}\right),\ \ \ Y_{3d} = -\sqrt{\frac{t_1}{2s_1} } \left(  \begin{array}{ccc} 0 & 0 & 0 \\ 1 & 0 & 0 \\ 1 & 0 & 0 \end{array}\ \right) 
	}[enilpotent]
These solutions are nilpotent (recall the discussion in Section \matrix); these vacua are inherently nondiagonalizable, with the D-terms satisfied via $[X,X^\dagger] = -[Y,Y^\dagger]$ for \enilpotent. Expanding \etoa in these vacua, the adjoints have mass terms, and so the low-energy theories are SQCD with extra massless flavors. The $\Delta W$ deformation in \etoa thus 
gives the RG flow
	\eqn{
	E_7\to A_2 + 5 A_1 + A_1^{2d} + A_1^{3d}(+\dots ?).
	}[etoavacuum]
The $(+\dots ?)$ indicate that there might be additional $d>3$ dimensional vacuum solutions, beyond the ones that we found here\footnote{We use the $SU(N)$ or $U(N)$ symmetry to gauge fix one real adjoint's worth of components in $X$, $X^\dagger$, $Y$, and $Y^\dagger$, and the remaining entries are constrained by the $D$- and $F$-terms, along with any Casimir conditions. We did not find an analytic way to construct, or exclude, higher-dimensional solutions beyond scanning computationally. Even gauge-fixing, scanning the solution space is harder for larger $d$, and so in \etoavacuum we only completed the scan for $d\leq 3$.}. In the following we will assume that there are no such additional vacua in \etoavacuum, but we do not have a proof that this is the case.  

If there are $n_0$ eigenvalues at the origin, $n_i$ in the $i$'th $A_1$ node, $n^{2d}$ in the $A_1^{2d}$ node, and $n^{3d}$ in the $A_1^{3d}$ node, such that $N_c=n_0+\sum_{i=1}^5 n_i + 2n^{2d} + 3n^{3d}$, then the gauge group is Higgsed in the electric and proposed magnetic descriptions (for $x$ in the conformal window):
	\eqna{
	\begin{array}{clc}
	U(N_c)  & \overset{\langle X \rangle, \langle Y \rangle}{\longrightarrow}\ & {\displaystyle U(n_0) \prod_{i=1}^{5} U(n_i) \times U(n^{2d}) \times U(n^{3d}) }\\ 
	\big\downarrow   & & \big\downarrow \\
	U(30N_f-N_c) &    \overset{\langle \hat{X} \rangle, \langle \hat{Y} \rangle}{\longrightarrow}\ & {\displaystyle U(2N_f - n_0) \prod_{i=1}^{5}U(N_f-n_i)\times U(2N_f - n^{2d}) \times U(3N_f - n^{3d})} \end{array}.
	}[EAflow]
The down arrows are Kutasov-Lin duality for the $E_7$ $U(N_c)$ theory in the UV, and Kutasov or Seiberg duality for the approximately decoupled low-energy gauge group factors in the IR. Comparing the UV and IR of the dual theories of the lower row of \EAflow as we did for the $D$-series, there is a mismatch in the dual gauge groups of $10N_f$.  Indeed, it is immediately evident that \sumofsquares\ is not satisfied for $\alpha =30$, since there are precisely 7 vacua with $d_i=1$, and $23\neq \sum d_i^2$ for integers $d_i>1$.  Something new is needed, beyond simply decoupled copies of SQCD in the various $d_i$-dimensional vacua. 

To recover the $SU(N_c)$ version of this flow, we must deform the superpotential \etoa by the operators $\text{Tr} Y^2,\ \text{Tr} XY,\ \text{Tr} X,\ \text{Tr} Y$, (the latter two with Lagrange multipliers) whose coefficients are shifted appropriately.  The 2d representation for the deformed superpotential smoothly matches onto the $U(N_c)$ solution in \enilpotent upon taking the coefficients of the lower order deformations to zero. The analogous check for the 3d representations in \enilpotent\ turns out to be technically challenging, and while we expect that it also matches, such that the  $SU(N_c)$ version of the flow will match onto \EAflow, we have not verified this. (For reasons that will become apparent in Section \ref{sec:dis}, this can be a subtle issue in the $E$-series.)

\subsubsec{$E_7\to D_5$: Disappearing vacua?}
\label{sec:dis}

We here consider the flow $E_7\to D_5$ for $U(N_c)$ gauge group, which corresponds to the superpotential (normalizing the couplings in the UV $E_7$ theory to one)
	\eqn{
	W = \text{Tr} \frac{1}{3} Y^3 +  \text{Tr} Y X^3 + t_1 \text{Tr}  X Y^2 + \frac{t_2}{4}  \text{Tr} X^4.
	}[edu]
The $F$-terms of \edu are given by
	\eqna{ 
	Y^2 + X^3 + t_1 \{X,Y\} = 0,\\
	Y X^2 + XYX + X^2 Y + t_1 Y^2 + t_2 X^3 = 0.
	}[edfterms]
The 1d vacuum structure along this flow consists of the $D_5$ theory at $X=Y=0$, and 2 $A_1$'s away from the origin. To study for higher-dimensional vacua we look for Casimirs, but
note that there are no simple Casimirs of \edfterms, except of course the $F$-terms \edfterms themselves. There is a 2d solution to the $F$-terms \edfterms and $D$-terms \Dterms of the form $X =  x_0 {\bf 1} + x_3 \sigma_3,\ Y = y_0 {\bf 1} + y_3 \sigma_3$, where $\{ x_0,x_3,y_0,y_3\}$ are determined functions of the couplings $t_1$ and $t_2$. Then, the RG flow leads to vacua
	\eqn{
	U(N_c),\quad t_i\ \text{generic}:\quad E_7\to D_5 + 2A_1 + A_1^{2d} \ (+\dots ?).
	}[etodvacuum]
As in \etoavacuum\ and the associated footnote, there might additionally be $d>3$ vacua, indicated here with $(+\dots ?)$. Performing the same check as in \EAflow, assuming Kutasov-Lin duality for the UV  $U(N_c)$ $E_7$ theory, there is a mismatch in the UV and IR dual gauge groups, this time of $15N_f$:
	\eqna{
	\begin{array}{clc}
	U(N_c)  & \overset{\langle X \rangle, \langle Y \rangle}{\longrightarrow}\ & {\displaystyle U(n_0) \prod_{i=1}^{2} U(n_i) \times U(n^{2d})}\\ 
	\big\downarrow   & & \big\downarrow \\
	U(30N_f-N_c) &    \overset{\langle \hat{X} \rangle, \langle \hat{Y} \rangle}{\longrightarrow}\ & {\displaystyle U(9N_f - n_0) \prod_{i=1}^{2}U(N_f-n_i)\times U(2N_f - n^{2d}) } 	\end{array}.
	}[EDflow]
	
The flow \etodvacuum\ is for generic deformations $t_1,\ t_2$  in \edu. There are special values of the coupling $t_2$ for which the 2d representation ``goes away" because $X$ or $Y$ becomes proportional to the identity, or proportional to each other---in either case, the solution is then accounted for by 1d vacua.  This possibility does not occur for the $D_{k+2}$ RG flows. The resulting flows are summarized by (for the rest of this subsection we refrain from putting the (+\dots ?), but note that everywhere there is the possibility of $d>2$ dimensional vacua):
	\eqna{
U(N_c),\quad t_2 = t_1 (7 \pm 2\sqrt{6})\ \ \text{or}\ \ 5t_1\frac{(-6\mp \sqrt{6})}{(-6\pm \sqrt{6})}:\quad  E_7 \to D_5 + A_2
	}[etodvactwo]
	\eqna{
U(N_c),\quad\quad t_2=t_1\ \ \text{or}\ \ t_1(1\pm \sqrt{6}):\quad\quad  E_7 \to D_5 + 2 A_1
	}[etwodvacthree]
For the flow \etodvactwo, the eigenvalues corresponding to the 1d and 2d $A_1$ singularities in \etodvacuum\ come together, enhancing to an $A_2$ singularity. Labeling the multiplicities of $X$ and $Y$'s eigenvalues as in \EDflow, then for the enhancement \etodvactwo the eigenvalues rearrange such that the electric theory is Higgsed $U(N_c)  \to U(n_0) \times U(n_1 + n_2 + 2n^{2d})$. For the case \etwodvacthree, the eigenvalues corresponding to the $A_1^{2d}$ theory in \etodvacuum\ match onto copies of the eigenvalues corresponding to the 1d $A_1$ theories, such that in the IR the vacua are $D_5 + 2 A_1$. In this case, the eigenvalues in the electric version of the flow rearrange such that $U(N_c)  \to U(n_0) \times U(n_1 + n^{2d})\times U(n_2 + n^{2d})$. 

This feature that a 2d representation can ``go away" is also present in the $SU(N_c)$ version of the flow \edu. As was the case for the $D$-series flows discussed in Section \ref{sec:Dshift}, there are multiple sets of deformations $\Delta W$ that one can add to \edu to recover the same 1d vacuum structure as in \etodvacuum for $SU(N_c)$ gauge group\foot{There are at least three possible sets of deformations, and we've explicitly checked that two of these (including (\ref{eq:defposs})) yield the same 2d vacuum structure.}. For instance, one possibility  is
	\eqna{
	\Delta W =  v_1  \text{Tr} X^2 Y + \frac{v_2}{3}  \text{Tr} X^3 + v_3  \text{Tr} XY + v_4  \text{Tr} X^2 + v_5  \text{Tr} Y^2 - \lambda_x  \text{Tr} X - \lambda_y  \text{Tr} Y. \label{eq:defposs}
	}[]
 Surprisingly, there are three 2d vacua for the flow \edu plus, for instance, (\ref{eq:defposs}): one which matches continuously onto the $U(N_c)$ 2d vacuum when the couplings of the lower order deformations are taken to zero, and two which do not. The additional two 2d vacua have the property that $X$ and $Y$ become proportional to each other in the limit that the couplings of the additional deformations (e.g., the $v_i$ and $\lambda_x,\lambda_y$ in (\ref{eq:defposs})) vanish. In other words, these additional vacua vanish precisely when we cannot perform the shift of the $SU(N_c)$ flow to the preferred $U(N_c)$ origin---i.e., when we can only flow down to decoupled $A_1$ theories in the IR. In sum, the vacua of this flow are 
 	\eqn{
	SU(N_c),\quad t_i\ \text{generic}:\quad E_7\to D_5 + 2A_1 + 3A_1^{2d}.
	}[suetod]
 However, as with our similar previous examples, using the known duals of the IR theories in \suetod\ does not fit with the $\alpha _{E_7}=30$ of  Kutasov-Lin duality, essentially because \sumofsquares\ is not satisfied: here it is because $\alpha _{E_7}\neq \alpha _{D_5}+2\alpha _{A_1}+3 \times 2^2 \alpha _{A_1}$, i.e. $30\neq 9+2+12$.

Analogously to the $U(N_c)$ flow \etodvacuum, one of the 2d vacua in \suetod reduces to 1d vacua in special cases. The difference here is that the other two 2d vacua in \suetod remain:  
	\eqna{
SU(N_c),\quad t_2 = t_1 (7 \pm 2\sqrt{6}):\quad  E_7 \to D_5 + A_2+ 2 A_1^{2d}
	}[etodvactwo]
	\eqna{
SU(N_c),\quad t_2=t_1\ \ \text{or}\ \ t_1(1\pm \sqrt{6}):\quad  E_7 \to D_5 + 2 A_1+ 2 A_1^{2d}
	}[etwodvacthree]
This feature that the 2d vacua can ``disappear" for particular values of the couplings is reminiscent of the wall crossing phenomena for BPS states. There are hints that this is a general phenomenon in the $E$-series. For instance, there is a similar effect in the $E_8\to D_6$ flow, as we discuss in Appendix \ref{sec:thirdcase}. It is presently unclear to us how to this phenomenon fits with proposed duals,  and we leave such an exploration for future work.

\subsubsec{$E_7\to A_6$: A seemingly irrelevant deformation}[thirdflow]

As expected from Arnold's singularities and deformations, there are RG flows corresponding to adjoint Higgsing of $G=A,D,E$.  For some of these Higgsing patterns, the corresponding $\Delta W$ deformation is not  immediately apparent. A general treatment of how to deform and resolve the ADE singularities by giving expectations values to the Cartan elements is described in \rcite{MR1158626}, and this formalism is applied in \rcite{Katz:1996xe} to several of the resolutions of present interest to us. We used an adaption of this construction to obtain the deformations of this section. 

We here consider the $\Delta W$ deformation which leads to the RG flow $E_7\to A_6$.  This is given $\Delta  W\sim \text{Tr} X^7$.  At first glance, this $\Delta W$ seems irrelevant at the $W_{E_7}$ SCFT, since it scales with a higher $U(1)_R$ charge than the terms in \Esup, but we know that such a flow should be possible (for instance, we can cut the $E_7$ Dynkin diagram to recover the $A_6$ diagram, as demonstrated for other cases in Figure \ref{fig:ahiggsing}). The resolution to this puzzle is that only a special shift of the deformation couplings will recover the $A_6$ singularity in the IR---even for the $U(N_c)$ case. The clearest way to see the enhancement of the $A_6$ singularity is through a change of variables. Since the change of variables is already complicated in the $U(N_c)$ case, we will only consider this flow for $U(N_c)$ gauge group here. We analyze other $E$-series flows whose $\Delta W$ deformations seem irrelevant in Appendix \ref{sec:irr}.

We start with $W_{E_7}$ plus $\Delta W$ deformations,
	\eqn{
	W = \frac{1}{3} \text{Tr}Y^3 + s_1  \text{Tr}Y X^3 + t_1  \text{Tr}X Y^2 + \frac{T_2}{2} \text{Tr} Y^2 + T_3 \text{Tr} X Y + \frac{T_4}{2} \text{Tr} X^2.
	}[start]
It follows from the $F$-terms of \start that there are seven 1d vacua in the IR, corresponding to seven $A_1$ theories (we will discuss higher-dimensional vacua below). It is useful to next linearly shift the fields $X\to X+n,\ Y\to Y+m$, where we choose $m$ and $n$ as functions of the couplings in \start to cancel in linear terms in $X$ and $Y$ which result from the change of variables. Dropping constants, the superpotential can then be rewritten as	
	\eqn{
	W = \frac{1}{3} \text{Tr}Y^3 + s_1  \text{Tr}Y X^3 + t_1  \text{Tr}X Y^2 + t_2 Y X^2 + \frac{t_3}{3}\text{Tr} X^3 + \frac{t_4}{2} \text{Tr}Y^2 + t_5 \text{Tr}XY + \frac{t_6}{2} \text{Tr} X^2,
	}[two]
where the $t_i$'s are defined in terms of the couplings in \start and $m,n$. We then implement the following change of variables for all $t_1\neq 0$:
	\eqn{
	Y = U - \frac{3t_1}{7} X - \frac{7s_1}{t_1} X^2 - \frac{343s_1^2}{96t_1^3}X^3.
	}[secondchange]
Such a change of variables is holomorphic, and has the property that the new field variable $U$ is single-valued in terms of the variable being replaced ($Y$). \secondchange shifts around the R-charges of the fields, but causes no problems; in particular, the metric in the scalar potential acts to compensate and keep the actual vacua the same. Rewriting \two in terms of $U$ and $X$ will result in many terms, including the terms $\text{Tr}X^7 $ and $\text{Tr} U^2$ which we identify as corresponding to the $A_6$ theory and which are now apparently relevant from the perspective of the UV theory, plus eight even more relevant deformations. 

So far, all we've accomplished is to rewrite the flow $E_7\to 7A_1$ in a complicated way. At this point, however, one can show that there is a unique shift of the couplings $\{t_2,t_3,t_4,t_5,t_6\}$ in terms of $t_1,s_1$, such that all of the coefficients to terms more relevant than those which we will identify with the $A_6$ theory vanish. Implementing this shift, \two becomes
	\eqna{
	W&= \frac{1}{3} \text{Tr}  U^3  -\frac{343 s_1^2}{96 t_1^3} \text{Tr} U^2 X^3 + \frac{117649 s_1^4}{9216 t_1^6} \text{Tr} U X^6-\frac{40353607 s_1^6}{2654208 t_1^9} \text{Tr} X^9-\frac{7s_1}{4t_1} \text{Tr} U^2 X^2 \\
	&+\frac{2401 s_1^3}{192 t_1^4}\text{Tr}  U X^5 - \frac{823543 s_1^5}{36864 t_1^7} \text{Tr} X^8 +\frac{4t_1}{7} \text{Tr} U^2 X -\frac{49 s_1^2 }{48 t_1^2} \text{Tr} U X^4 - \frac{16807 s_1^4}{4608 t_1^5} \text{Tr} X^7 \\
	& - \frac{48 t_1^3}{343s_1} \text{Tr} U^2.
	}[final]
Studying the $F$-terms of this superpotential and expanding \final in the vacua, there is one vacuum at the origin corresponding to the $A_6$ theory, and one away from the origin corresponding to an $A_1$ theory. Thus, we have recovered the desired flow.

We have also studied the 2d vacuum structure of this RG flow\foot{We have not as of writing attempted to find $d>2$ dimensional vacua for this flow.}. For generic values of the couplings in \two, there are nine 2d vacua which we can parameterize as $X = x_0{\bf 1} + x_3\sigma_3,\ Y = y_0 {\bf 1} + y_3 \sigma_3$, such that the generic $\Delta W$ deformations lead to the vacua
	\eqn{
	E_7\to 7A_1 + 9A_1^{2d}.
	}[etoageneric]
However, all of these 2d vacua ``go away" in the enhancement to the $A_6$ theory, in the sense described in Section \ref{sec:dis}. In particular, of the 18 eigenvalue pairs corresponding to the $A_1^{2d}$'s in \etoageneric, 15 come to the origin to form the $A_6$ theory in the shift to \final, while the remaining 3 become copies of the shifted $A_1$ theory. Thus the 1 and 2d vacuum structure of this flow appears to be
	\eqn{E_7\to A_6 + A_1}[etoaspecial]
where the multiplicities of the eigenvalues corresponding to the higher-dimensional vacua of \etoageneric have redistributed appropriately.


\newsec{Conclusions, future directions, and open questions}\label{sec:E68flows}

\subsec{Recap: some puzzles and open questions for the $D_\text{even}$ and $E_7$ theories}

The ADE SCFTs have a rich structure of vacua, and deformations.  The fact that the fields $X$ and $Y$ are matrices introduces many novelties, as we have here illustrated---but not yet fully understood. It is natural to expect that the higher-dimensional representations of the $F$- and $D$-terms have dimensions $d_i$ given by some $G=A,D,E$ group theory quantities, e.g. the Dynkin indices $n_i$ as with the McKay correspondence.  But we find that $d_i\neq n_i$  in general, and we do not yet know how to analytically find the $d_i$ and associated representations. 

 Our analysis of the $E$-series shows that even associating a fixed set of representations with the deformation flow can be subtle.  For example, the case studies of Section \ref{sec:Eflows} give the following puzzle: we can RG flow from the $W_{E_7}$ SCFT via different $\Delta W$ deformations, to decoupled copies of SQCD ($A_1$) at low energies, and for different routes seemingly get different numbers of higher-dimensional representations in the IR. It will be interesting to understand how the proposed duality \rcite{DKJLyqa} fits in with this picture. The present  work has raised several additional hurdles for the conjectured $D_{\text{even}}$ and $E_7$ dualities, and it will be interesting to see how all of these puzzles are resolved.

\subsec{Future directions: aspects and challenges of the $W_{E_6}$ and $W_{E_8}$ theories}\label{sec:Eseriesdisc}

The superpotentials that drive the RG flow from $\hat O\to \hat E\to E_{6,8}$ are
\matrixArnold:
	\eqn{
	W_{E_6} = \frac{1}{3} \text{Tr} Y^3 + \frac{s}{4}\text{Tr}  X^4.
	}[Ebsup]
	\eqn{
	W_{E_8} = \frac{1}{3} \text{Tr} Y^3 + \frac{s}{5}\text{Tr}  X^5.
	}[Ecsup]
The $\tr X^4$ and $\tr X^5$ terms are relevant for $x^{E_6}_{\text{min}}\approx 2.44$ and $x^{E_8}_{\text{min}}\approx 7.28$, respectively \rcite{twoadj}.  As reviewed in Section \chiralringmatrix, the chiral rings of these theories do not classically truncate, and are especially rich since $X$ and $Y$ decouple in the $F$-terms \Esixfterm\ and \Eeightfterm.  As shown in \rcite{DKJLyqa,Kutasov:2014wwa}, the $W_{E_{6,8}}$ theories cannot have a dual of the form reviewed following \Adualalpha. It is unknown if there is a dual of some different form.

The $a$-theorem condition \aderivneg\ is violated for sufficiently large $x$ for both theories \rcite{twoadj}, showing that some new quantum effects must arise for large $x$.  One possibility is that a $W_{dyn}$ is generated, and the theory is no-longer conformal, for some $x>x_\text{stability}$. Another possibility is that there is some unknown dual description which becomes IR-free for large $x$.  There are other reasons to expect that there might be some description of the IR physics of \Ebsup and \Ecsup in terms of dual variables: we can flow, for instance, $E_6\to D_5$, and we expect that the stability bound is reduced $x^{\text{max}}_{E_6}>x^{\text{max}}_{D_5}$ along RG flow. It is also pointed out in \rcite{DKJLyqa,Kutasov:2014wwa} that in $E_6$ the number of operators at a given value of $R$ grows with R-charge, but somehow the theory must find a way to preserve unitarity. 


We have studied a few aspects of the moduli space and $\Delta W$ deformations of the  $W_{E_6}$ and $W_{E_8}$ SCFTs, looking for clues in formulating a dual description of the theories, but finding puzzles (similar to $D_\text{even}$ and $W_{E_7}$). We here briefly report on some of our findings.

The undeformed $W_{E_6}$ and $W_{E_8}$ theories have a variety of flat directions similar to those discussed for the $W_{D_{\text{even}}}$ and $W_{E_7}$ theories in Sections \dline and \ematrix. In particular, both have 2d and 3d nilpotent flat directions (of course, a flat direction of $E_6$ is also a flat direction of $E_8$, since $X^3=0\Rightarrow X^4=0$). The 2d vacuum solutions are of the form $ X_{2d}= x(\sigma_3 + i \sigma_1),\ Y_{2d} = - x(i\sigma_3 + \sigma_1)$ where arbitrary complex $x$ labels the flat direction. There are several 3d flat directions of these theories, again labeled by $x$, for instance
	\eqna{
	X_{3d} = x \left(\begin{array}{ccc} 0 & 0 & 1 \\ 0 & 0 & 0 \\ 0 & 1 & 0 \end{array}\right),\ \ Y_{3d}  =  x \left(\begin{array}{ccc} 0 & 0 & 0 \\ 1 & 0 & 0 \\ 0 & 0 & 0 \end{array}\right),\\
	X_{3d'} =  x \left(\begin{array}{ccc} 0 & 1 & 1 \\ 0 & 0 & 0 \\ 0 & 0 & 0 \end{array}\right),\ \ Y_{3d'} = x  \left(\begin{array}{ccc} 0 & 0 & 0 \\ 1 & 0 & 0 \\ 1 & 0 & 0 \end{array}\right).
	}[threedfd]
As with the $D_{\text{even}}$ and $E_7$ cases, these (classical) flat directions are surely related to the classical nontruncation of the ring.  We expect, as with those cases, that some dynamics must alter these flat directions, at least for sufficiently large $x$, to avoid apparent violations of the $a$-theorem. It would be interesting to understand this further. 

For $SU(N_c)$, as opposed to $U(N_c)$,  upon imposing the tracelessness of the adjoints by adding Lagrange multiplier terms to \Ebsup and \Ecsup, these theories have $SU(N_c)$ flat directions for particular values of $N_c$, similar to those discussed in Section \FDAk, \dsun, and \esun. The $W_{E_6}$ theory has a flat direction for $N_c=3m$ and/or $N_c=2n$, while $E_8$ has a flat direction for $N_c = 2n$, for integer $m$ and $n$. We expect low-energy $\widehat A$ theories along these classical flat directions; it would be interesting if one can obtain insights about the theory at the origin from these flat directions. 


We now briefly comment on the RG flows from some $\Delta W$ deformations of the $W_{E_6}$ and $W_{E_8}$ SCFTs.  Consider e.g. the flow $E_6\to D_5$, obtained via adding $\Delta W=\tr X Y^2$ to \Ebsup. The 1d vacua correspond to the $D_5$ theory at the origin, and an $A_1$ theory away from the origin. The $F$-terms imply that $[Y^2,X]=0$, and $[X^2,Y]=[X^3,Y]=0$, so that $d>1$ dimensional solutions to the $F$-terms must actually satisfy $X^2=0$. It is then straightforward to show that there are no 2d or 3d solutions that satisfy the $F$-terms and $D$-terms, so that the vacua along the flow are just the 1d vacua (up to possible $d>3$ representations, again as in the discussion around \etoavacuum)
	\eqn{
	E_6\to D_5 +  A_1 \qquad (+\dots?).
	}[esixtodfive]
While we do not yet know of a dual description of the $W_{E_6}$ SCFT, in the IR of this flow Brodie duality and Seiberg duality map the low-energy gauge groups as 
	\eqn{
	U(n_0) \times U(n_1)\ \overset{\text{duality}}{\longrightarrow} U(9N_f-n_0)\times U(N_f-n_1),\quad N_c=n_0+n_1.
	}[esixtodfivehiggs]
Perhaps understanding the IR limits of such flows will yield hints pointing towards a dual description of the $W_{E_6}$,$W_{E_8}$ theories.  We invite the interested reader to try. Some additional comments on $E$-series flows are provided in Appendix \ref{sec:irr}.

\ack{We would like to thank David Kutasov, Sergei Gukov, Jennifer Lin, Luca Mazzucato (who also shared some unpublished notes on related topics), John McGreevy, and Nathan Seiberg for helpful discussions. This work was supported by DOE grant  DE-SC0009919.}


\begin{appendices}


\newsec{Chiral ring elements of $E_7$}  \label{sec:cr}

We list the 30 independent generators $\Theta_j,\ j=1,...,30$ of the $E_7$ chiral ring in \rcite{DKJLyqa}, where $N$ is the polynomial degree. 

\begin{center}
\begin{tabular}[c]{l l l | l l l}
$j$ & $N$ & $\Theta_j$ &$j$ & $N$ & $\Theta_j$\\
\hline 
1 & 1 & 1 & 16 & 11& $YX^4$\\
2 & 2 & $X$ & 17 & 11& $XYX^3$\\
3 & 3 & $Y$ &18 & 12& $X^6$\\
4 & 4  & $X^2$ & 19 & 12& $YXYX^2$\\
5 & 5 & $YX$ & 20 & 13& $YX^5$\\
6 & 5& $XY$ & 21 & 13& $XYX^4$\\
7 &6 & $X^3$ & 22 & 14& $X^7$\\
8 & 7& $YX^2$ & 23 & 14& $YXYX^3$\\
9 & 7& $XYX$ & 24 & 15& $YX^6$\\
10 & 8& $X^4$ & 25 & 16& $X^8$\\
11& 8& $YXY$ & 26 & 16& $YXYX^4$\\
12 &9& $YX^3$ & 27 & 17& $YX^7$\\
13 & 9& $XYX^2$ & 28 & 18& $X^9$\\
14 &10 & $X^5$ & 29 & 19& $YX^8$\\
15 & 10& $YXYX$ & 30 & 21 & $YX^9$
\label{table:magnetic}
\end{tabular}
\end{center}


\newsec{RG Flows whose deformations seem irrelevant}  \label{sec:irr}

We briefly consider (as in Section \thirdflow) some cases where the $\Delta W$s, corresponding to some ADE adjoint Higgsing pattern, are not immediately apparent.  We focus on recovering the desired 1d vacuum structure for $U(N_c)$ flows, leaving a full analysis of the higher-dimensional structure for future work. The cases studied in Sections \ref{sec:firstcase} and \ref{sec:secondcase} are analogous to singularity resolutions studied in \rcite{Katz:1996xe}.

\subsection{$E_6\rightarrow A_5$}\label{sec:firstcase}

We start with the deformed $E_6$ superpotential,
	\eqn{
	W= \frac{1}{3} \tr Y^3 + \frac{s}{4} \tr X^4 + t_1 \tr Y X^2 + t_2 \tr  Y^2,
	}[edsup]
whose $F$-terms are
	\eqna{
	Y^2 + t_1 X^2 + 2 t_2 Y &= 0 \\
	s X^3 + t_1 \{X,Y\} &= 0.
	}[ftermsesix]
For 1d representations, $X=x{\bf 1},Y=y{\bf 1}$, \ftermsesix yield vacua which correspond to the following IR theories (as usual, seen by expanding \edsup in each vacuum):
	\eqna{
(x,y) = \left\{ \begin{array}{ccc}  (0,0) & \leftrightarrow & A_3 \\
 (0,-2t_2)& \leftrightarrow & A_1\\
\left(\pm\sqrt{ \frac{4t_1}{s} \left(t_2 - \frac{t_1^2}{s}\right)}, -2 \left(t_2 - \frac{t_1^2}{s}\right)\right) & \leftrightarrow  & 2A_1\end{array}\right.
}[eigsesix]

For the special value of $t_2 = \frac{t_1^2}{s}$, for nonzero $s$, the two eigenvalues on the last line of \eigsesix collapse to the origin to enhance the $A_3$ singularity. This is more clearly seen by changing variables $Y=\frac{s}{t_1}(Z-X^2)$. Then, for the special value of  $t_2 = \frac{t_1^2}{s}$, \edsup rewritten in terms of the $X,Z$ fields gives the $A_5$ theory at the origin from the $\tr X^6$ term in 
	\eqn{
	W=  \frac{s^3}{t_1^3}  \left( \frac{1}{3}  \tr Z^3 -  \tr Z^2 X^2 +  \tr Z X^4\right) - \frac{s^3}{3t_1^3}  \tr X^6 + s \tr Z^2.
	}[xz]
The $F$-terms of \xz then yields the 1d vacua $A_5 + A_1$. 

To find higher dimensional representations of vacua for this flow, we note that \ftermsesix implies $[X^2,Y]=0$.  Since $[Y^2,X] \backsim [X,Y]$, $Y^2$ is not a Casimir; instead we use $Y^2 + 2 t_2 Y = f(y) {\bf 1}$. Parameterizing 2d solutions by $Y = y_0 {\bf 1} + y_1 \sigma_1 $ and $X = x \sigma_3$, $Y$'s Casimir condition fixes $y_0$, so that the $F$-terms for 2d vacua simplify to
	\eqna{
	(-t_2^2 + y_1^2 + t_1 x^2){\bf 1} &= 0 \\
	x(s x^2  - 2 t_1 t_2) \sigma_3 &= 0.
	}[twodfterms]
The second relation in \twodfterms fixes the eigenvalue $x$, and the first fixes $y_1$, such that we indeed have a 2-dimensional vacuum (only one, as gauge symmetry relates $x\to -x$ and $y_1\to - y_1$). This vacuum exists both for generic $t_2$, and for $t_2$ shifted to give the $A_5$ theory. In sum, the flow \edsup has the following 1d and 2d vacua:
	\eqna{
	t_2\ \text{generic}:\quad E_6 &\to A_3 + 3 A_1 + A_1^{2d}\ (+\dots?) \\
	t_2=\frac{t_1^2}{s}: \quad E_6 &\to A_5 + A_1 + A_1^{2d}\ (+\dots?).
	}[esixtoaflow]

\subsection{$E_7\rightarrow D_6$}\label{sec:secondcase}

Here, we start with the $E_7$ superpotential deformed by the $D$-series term $\tr XY^2$,
	\eqn{
	W_{E_7}+\Delta W = \frac{1}{3} \tr Y^3 + s \tr Y X^3 + t \tr X Y^2.
	}[edsupb]
There are two sets of 1-dimensional vacuum solutions for $X$ and $Y$, corresponding to the eigenvalues $(x=0,y=0)$, and $(x=\frac{5 t^2}{9s}, y=-\frac{25 t^3}{27s})$. Expanding near the origin 
appears to just give $W_{low} \backsim \tr X Y^2=W_{\widehat D}$. Consider though the following sequence of variable changes:
	\eqna{
	X =U - \frac{1}{3 t}Y,\quad Y = \frac{s}{2 t} (Z-U^2).
	}[ch]
In terms of the $U,Z$ fields, \edsupb becomes  
	\eqna{
	W=   & \frac{s^5}{108 t^7}  \left(  - \frac{1}{4} \tr U^8 + \tr Z U^6  - \tr Z^2 U^4  -  \frac{1}{2}\tr (ZU^2)^2 + \tr Z^3 U^2   - \frac{1}{4} \tr Z^4 \right)   \\ 
	+ &  \frac{s^4}{24 t^5}  \bigg(- \tr U^7   + 3 \tr ZU^5  - 2 \tr Z^2U^3 - \tr U Z U^2 Z  + \tr Z^3U    \bigg)  \\
	+ &  \frac{s^3}{4 t^3}   \left( -  \tr U^6  + 2 \tr ZU^4  -  \frac{2}{3} \tr Z^2 U^2-  \frac{1}{3} \tr (UZ)^2\right) +\frac{s^2}{4 t}\bigg(\tr U Z^2  -\tr  U^5 \bigg).
	}[uz]
We've organized the terms in \uz by increasing relevance from the perspective of the UV fixed point. The most relevant terms in the IR limit of the flow are those in the last parentheses, such that the $D_6$ theory resides at the origin. There is a 1d vacuum solution to the $F$-terms of \uz corresponding to an $A_1$ theory, such for all $t\neq 0$ that we recover the 1d vacua:
	\eqna{
	E_7\rightarrow D_6+A_1\ (+\dots?)}[eseventodsix]
where here the  (+\dots?) refers to all $d>1$ dimensional vacuum solutions.

\subsection{$E_8\rightarrow D_7$} \label{sec:thirdcase}

We start by deforming the $E_8$ theory with a $D$-series deformation and $E_7$ deformation, 
	\eqn{
	W =  \frac{1}{3} \tr Y^3 + \frac{s}{5} \tr X^5 + t_1 \tr  X^3 Y + t_2 \tr X Y^2.
	}[supee]
From the 1d $F$-terms of this superpotential, there is eigenvalue pair at the origin  and two away from the origin. As in the previous subsection, there is naively some ambiguity in identifying the solution at the origin, since each of $\text{Tr}X^3Y$, $\text{Tr} X^5$, and $\text{Tr} X Y^2$ appear to be marginal deformations of the UV theory, but the eigenvalue decomposition suggests that the theory at the origin corresponds to a $D_6$. Then, the 1d vacua of \supee are $D_6+2A_1$.

There is a particular shift of the coefficients $t_2= \frac{5 t_1^2}{4 s} \equiv t_*$ that brings one of the nonzero $A_1$ eigenvalue pairs to the origin. A change of variables clarifies what is happening: take $Y=U-\frac{2s }{5t_1}X^2$, such that \supee becomes 
	\eqna{
	W&=\frac{1}{3} \tr U^3 - \frac{2 s}{5t_1}\tr  U^2 X^2 + \frac{4 s^2}{25 t_1^2} \tr  U X^4 - \frac{8 s^3}{375 t_1^3} \tr  X^6+ \left(t_1 - \frac{4 s t_2}{5t_1}\right) \tr U X^3  \\
	&-\frac{s}{5t_1} \left(t_1-\frac{4 s t_2}{5 t_1}  \right) \tr X^5+ \frac{5 t_1^2}{4 s} \tr U^2 X.
	}[estep]
The 1d $F$-terms of \estep still yield one zero eigenvalue pair and two nonzero eigenvalue pairs, but if we now shift $t_2=t_*$, then the $D_6$ theory at the origin is enhanced to a $D_7$ theory, while only one nonzero (1d) vacuum remains, in which both $X$ and $U$ receive masses. In some, the shift $t_2=t_*$ results in the 1d vacua $D_7 + A_1$.  

We now study higher-dimensional representations of vacuum solutions to the $F$-terms of \supee and $D$-terms \Dterms. For generic values of the couplings, there is a 2d vacuum (letting $s=1$) 
	\eqna{
	X &= x_0 {\bf 1} + x_3 \sigma_3,\quad Y=y_0 {\bf 1} + y_3 \sigma_3,\\
	x_0 &= t_1 (-\frac{9}{2} t_1^2 + 4 t_2),\quad x_3 = \frac{1}{2} (9 t_1^2 - 4 t_2)^{1/2} (3 t_1^2 - 2 t_2),\\ y_0&=\frac{1}{2} t_1 (-27 t_1^4 + 45 t_1^2 t_2 - 20 t_2^2),\quad y_3 = \frac{1}{2} (9 t_1^2 - 4 t_2)^{3/2} (t_1^2 - t_2).
	}[thetwodvacuum]
Then, for generic values of $t_1$ and $t_2$, the 1d and 2d vacua of this flow are
	\eqna{
	E_8 \to D_6 + 2 A_1 + A_1^{2d}\ (+\dots?).
	}[eeightflow]
As is evident in \thetwodvacuum, there exist special values of $t_2$ for which the 2d vacua ``go away" in the sense of Section \ref{sec:dis}, e.g. 
	\eqna{
	t_2=\frac{5t_1^2}{4}\equiv t_*:\quad 	E_8 \to D_7 + A_1\ (+\dots ?)
	}[eighta]
	\eqna{
	t_2=\frac{3t_1^2}{2}\ \ \text{or}\ \ t_1^2:\quad E_8 \to D_6 + 2A_1\ (+\dots ?)
	}[eightb]
	\eqna{
	t_2=\frac{9t_1^2}{4}:\quad\quad	E_8 \to D_6 + A_2\ (+\dots ?)
	}[eightc]
In all cases above, the (+\dots ?) refers to $d>2$ dimensional vacua. The special case \eighta corresponds precisely to the shift $t_*$ already discussed, in which the $D_6$ singularity is enhanced to a $D_7$ singularity. In this case, one of the two eigenvalues corresponding to an $A_1^{2d}$ in \thetwodvacuum\ goes the origin, and the other becomes a copy of the eigenvalues corresponding to the remaining $A_1$ theory. In \eightb, the eigenvalues corresponding to the $A_1^{2d}$ theories in \thetwodvacuum become copies of the eigenvalues corresponding to the 1d $A_1$ theories. For the shift in \eightc, the two $A_1$ theories as well as the $A_1^{2d}$ theory in \thetwodvacuum are enhanced to an $A_2$ theory.

\end{appendices}

\bibliography{ADEdraftfinal}{}

\end{document}